\documentclass[fleqn,usenatbib]{mnras}

\usepackage{newtxtext,newtxmath}

\usepackage[T1]{fontenc}

\DeclareRobustCommand{\VAN}[3]{#2}
\let\VANthebibliography\thebibliography
\def\thebibliography{\DeclareRobustCommand{\VAN}[3]{##3}\VANthebibliography}

\usepackage{graphicx}	
\usepackage{amsmath}	

\usepackage{hyperref}
\hypersetup{colorlinks=true,allcolors=blue,pdfborder={0 0 0}}

\title[ALMA millimetre-wavelength imaging of HD 138965]{ALMA millimetre-wavelength imaging of HD 138965: New constraints on the debris dust composition and presence of planetary companions}

\author[J. P. Marshall et al.]{
J. P. Marshall,$^{1}$\thanks{E-mail: jmarshall@asiaa.sinica.edu.tw (JPM)}
S. Hengst,$^{2}$
A. Trejo-Cruz,$^{3}$
C. del Burgo,$^{4,5}$
J. Milli,$^{6}$
M. Booth,$^{7}$
J.C. Augereau,$^{8}$
\newauthor
E. Choquet,$^{9}$ 
F. Y. Morales,$^{10,11}$ 
P. Th\'ebault,$^{12}$ 
F. Kemper,$^{13,14,15}$ 
V. Faramaz-Gorka$^{16}$ and 
G. Bryden$^{10}$
\\
$^{1}$Institute of Astronomy and Astrophysics, Academia Sinica, 11F of AS/NTU Astronomy-Mathematics Building, No.1, Sec. 4, Roosevelt Rd, Taipei 106319, Taiwan\\
$^{2}$School of Mathematics, Physics, and Computing, University of Southern Queensland, West Street, Toowoomba, QLD 4350, Australia\\
$^{3}$Instituto de Radioastronom\'ia y Astrof\'isica, Universidad Nacional Aut\'onoma de M\'exico, Antigua Carretera a P\'atzcuaro \# 8701, Ex-Hda. San Jos\'e de la Huerta, \\Morelia, Michoac\'an, M\'exico C.P. 58089\\
$^{4}$Instituto de Astrof\'\i sica de Canarias, V\'\i a L\'actea S/N, La Laguna 38200, Tenerife, Spain\\
$^{5}$Departamento de Astrof\'\i sica, Universidad de la Laguna, La Laguna 38200, Tenerife, Spain\\
$^{6}$Institut de Plan\'etologie et d'Astrophysique de Grenoble, Universit\'e Grenoble Alpes, 621 Avenue Centrale, 38400 Saint-Martin-d'H\'eres, France\\
$^{7}$UK Astronomy Technology Centre, Royal Observatory Edinburgh, Blackford Hill, Edinburgh EH9 3HJ, UK\\
$^{8}$Universit\'e Grenoble Alpes, CNRS, IPAG, 38000 Grenoble, France\\
$^{9}$Aix-Marseille Universit\'e, CNRS, CNES, LAM, Marseille, France\\
$^{10}$Jet Propulsion Laboratory, California Institute of Technology, 4800 Oak Grove Drive, Pasadena, CA 91109, USA\\
$^{11}$STEM Division, Los Angeles Mission College, 13356 Eldridge Avenue, Sylmar, CA 91342, USA\\
$^{12}$LESIA-Observatoire de Paris, UPMC Univ. Paris 06, Univ. Paris-Diderot, France\\
$^{13}$Institut de Ciències de l'Espai (ICE, CSIC), Can Magrans, s/n, E-08193 Cerdanyola del Vallès, Barcelona, Spain\\
$^{14}$ICREA, Pg. Lluís Companys 23, E-08010 Barcelona, Spain\\
$^{15}$Institut d'Estudis Espacials de Catalunya (IEEC), E-08860 Castelldefels, Barcelona, Spain\\
$^{16}$Steward Observatory, Department of Astronomy, University of Arizona, 933 N. Cherry Ave, Tucson, AZ 85721, USA}
\date{Accepted XXX. Received YYY; in original form ZZZ}

\pubyear{2025}

\begin{document}
\label{firstpage}
\pagerange{\pageref{firstpage}--\pageref{lastpage}}
\maketitle

\begin{abstract}
HD~138965 is a young A type star and member of the nearby young Argus association. This star is surrounded by a broad, bright debris disc with two temperature components that was spatially resolved at far-infrared wavelengths by \textit{Herschel}. Here we present ALMA millimetre-wavelength imaging of the cool outer belt. These reveal its radial extent to be 150$^{+10}_{-7}$ au with a width ($\sigma$) of 49$^{+7}_{-6}$ au ($\Delta R/R = 0.77$), at a moderate inclination of 49\fdg9$^{+3.3}_{-3.7}$. Due to the limited angular resolution, signal-to-noise, and inclination we have no constraint on the disc's vertical scale height. We modelled the disc emission with both gravitational and radiation forces acting on the dust grains. As the inner belt has not been spatially resolved, we fixed its radius and width prior to modelling the outer belt. We find astronomical silicate is the best fit for the dust composition. However, we could not reject possible scenarios where there are at least 10\% water-ice inclusions. Combining the spatially resolved imaging by ALMA with non-detection at optical wavelengths by \textit{HST}, we obtain a limit on the scattering albedo $\omega \leq 0.09$ for the debris dust in the outer belt. Analysis of the outer belt's architecture in conjunction with simple stirring models places a mass limit of 2.3~$\pm$~0.4~$M_{\rm Jup}$ on a companion interior to the belt ($a \leq$ 78 au), a factor of two improvement over constraints from high contrast imaging.
\end{abstract}




\section{Introduction}


Main sequence stars play host to both planets and belts of planetesimals. Dynamical interaction between the components can excite the motions of the planetesimals to such a degree that a collisional cascade is triggered generating detectable amounts of dust. The dust produced in the collisional cascade is most commonly detected at optical or near-infrared wavelengths as scattered light \citep[in total intensity or polarisation, e.g.][]{2020Esposito,2023Ren,2023bMarshall,2024Crotts} or at infrared to millimetre wavelengths as thermal emission \citep[e.g.][]{2013Eiroa,2013Kennedy,2016Montesinos,2017Holland,2025Matra}. Due to their origins in destructive processes, these dusty structures are collectively referred to as `debris discs' \citep{2008Wyatt,2018Hughes}.

These debris discs are larger and more massive analogues to the Solar system's Asteroid or Edgeworth-Kuiper belts \citep{2020Horner}. There is some dependence of disc detection on the stellar luminosity, with the highest detection rate found around A stars \citep{2014Thureau}, an intermediate fraction around FGK stars \citep{2018Sibthorpe}, and the lowest detection rate around M dwarfs \citep{2009Lestrade,2018Kennedy}. However, these detections are strongly sensitivity limited at levels of fractional luminosity around ten times greater than that predicted for dust in the Solar system \citep[$\simeq 10^{-7}$,][]{2012Vitense}. As such, the supposed lack of cold debris discs around M dwarf stars could be a result of that limitation \citep{2020Luppe,2023Cronin}. Likewise, there is a dependence on disc detection with age. All else being equal, younger stars have brighter debris discs \citep{2007Wyatt,2011Kains}. This is consistent with debris discs originating from the steady collisional grinding of parent bodies into dust over the lifetime of the system \citep{2022Najita}. That said, the interstellar medium (ISM) can influence the morphology of debris discs throughout their life \citep{2025Heras}. As such, young stars with reliable ages (e.g. in moving groups) make favourable targets for the detection and characterisation of both debris discs and planetary companions.

HD~138965 is a young A star lying at a distance of 78.60~$\pm$~0.17~pc \citep{2016Gaia,2023aGaia,2023bGaia}. It has been identified as a member of the 40 to 50 Ma Argus association \citep{2019Zuckerman}, but an older age of 350$^{+37}_{-56}$ Ma based on fitting \textit{Gaia} observations has been also proposed \citep{2015Trevor}. The first evidence of cool infrared excess was obtained from \textit{IRAS} \citep{1991PattenWillson,2007Rhee}, with subsequent detections by \textit{Akari}/FIS \citep{2014Liu} and \textit{Spitzer} \citep{2009Morales,2014Chen}. The \textit{Spitzer}/IRS spectrum is featureless, but its shape is indicative that a substantial warm inner component to the debris disc is present within this system \citep{2014Chen}. The cool, outer disc was spatially resolved in \textit{Herschel}/PACS observations, with a measured radius of 187~$\pm$~6~au \citep{2016Morales}. 

The disc has yet to be imaged in scattered light, and remains one of the few relatively bright debris discs ($L_{\rm d}/L_{\star} > 10^{-4}$) around nearby stars without such observations. Combining multi-wavelength, spatially resolved imaging of debris disc systems with radiative transfer modelling is necessary to weaken degeneracies inherent between the dust properties and architecture \citep[e.g.][]{2014aMarshall}, thereby providing the greatest precision possible from the determination of the system's properties.

The architecture of HD~138965's debris disc, with two distinct temperature components, might imply the presence of two distinct planetesimal belts separated by (a) gap(s) produced by (a) planet(s), keeping the intervening region free of dust. Planetary companions influence the dynamics of planetesimal belts, stirring their motions and initiating the required collisional cascade and produce the observed debris disc \citep{2009Mustill}. Alternatively, the planetesimal belt may be stirred only by small bodies within the belt \citep{2018KrivovBooth}, or by some combination of these two factors \citep[e.g.][]{2023Munoz}. 

The radial and vertical structure of debris belts can potentially be used to constrain the mass of planetary companions stirring or sculpting them \citep[e.g.][]{2009Mustill,2018KrivovBooth,2019Daley,2021Marino}. Similarly, limits on the mass(es) of any potential companion(s) can be inferred from the overall architecture of the debris disc system \citep{2022Pearce} or from the analysis of the gas kinematics.  The power of constraints based on the disc architecture are limited by how well the belt is resolved and the nature of the interaction between companion and belt (i.e. stirring or sculpting). Existing imaging observations have not spatially resolved the width of the outer belt of HD~138965 \citep{2021Marshall}, so the constraints derived from such calculations remain weak. Theoretical models suggest that the extent of HD~138965's debris disc is consistent with self-stirring by planetesimals, independent of the assumed age of the system \citep{2018KrivovBooth}. The current best limits come from high contrast imaging observations of HD~138965 which rule out the presence of a 10~$M_{\rm J}$ companion beyond 70~au from the star, assuming a stellar age of $348^{+39}_{-54}$ Ma \citep{2018Matthews}. 

Here we present Atacama Large Millimiter/Submillimeter Array (ALMA) Band 6 (1.3~mm) observations of HD 138965's debris disc, for the first time spatially resolving the system at millimetre wavelengths. We interpret these observations in conjunction with the scattered light non-detection to place upper limits on the dust albedo through radiative transfer modelling. We further infer mass limits on planetary companions in the vicinity of the disc based on the belt architecture and stellar age. The remainder of the paper proceeds as follows. In Section \ref{sec:obs} we summarise the observations of HD~138965 that underpin the subsequent analysis. Then, in Section \ref{sec:mod} we present the modelling approach, for both the host star and the disc architecture, before presenting our results. Our findings are then discussed in context in Section \ref{sec:dis}. Finally, we summarise our results and present our conclusions in Section \ref{sec:con}.

\section{Observations}
\label{sec:obs}

Here we describe the various data sets obtained and combined to model HD~138965 and its accompanying debris disc. Most of these observations are obtained from publicly available catalogues or archive as science-ready, reduced products.

\subsection{ALMA}

Our target source HD~138965 (ICRS: RA = 15h 40m 11.395s, DEC = --70d 13m 41.47s) was observed by ALMA on 18 December, 2019, during its Cycle 6. The observations, under project code 2019.1.01220.S (PI: J.P. Marshall), were carried out using the 12 m array with a total of 42 antennas and employing a single pointing (no mosaic).

A total of four spectral windows (SPW) were used, including one centred at $\sim$ 230.1 GHz to target the CO (2-1) molecular line transition. The remaining three SPWs were centred at the topocentric frequencies of 227.2, 243.1, and 245.1 GHz and were employed to observe continuum emission. All SPWs fall in the ALMA Band 6 frequency
coverage. Observations are sensitive to emission with a largest angular scale of $\sim$ 12\farcs5, while the field of view is approximately a factor of two larger. The instrument setup chosen used a bandwidth of 2 and 1.875 GHz, for the continuum and line SPWs, respectively. The corresponding channel spacings are $\sim$ 15625 and 976.6 kHz.
Calibration and imaging of the single dataset were performed following standard procedures from the ALMA Observatory; during the observation, quasars J1427-4206 and J1558-6432 were used as bandpass/flux and phase/amplitude calibrators, respectively. The only dataset obtained for our ALMA program suffered from bad weather, resulting in amplitude drops for the flux calibrator during its observing scan. Affected data was removed, and a difference of at least 5\% in flux level was found before and after data flagging. This indicates a larger than expected flux uncertainty for the calibration, and therefore imaging, of HD 138965.

A total on-source time of 40 min was achieved for HD 138965. All imaging was done using a robust value of 0.5 in the Briggs weighting scheme. The final continuum image has a sensitivity of 3$\times$RMS $\sim70~\mu$Jy, with a restored (synthesised) beam of $1\farcs39~\times~1\farcs17$, at a position angle $\phi = 10\fdg5$. 
CO (2--1) line emission cubes were also inspected, with a channel width of $\sim$ 1.3 km s$^{-1}$ (RMS = 1.5 mJy beam$^{-1}$) and binned to 10 km s$^{-1}$ (RMS = 0.67 mJy beam$^{-1}$). Detailed data analysis and properties of the continuum and CO line emission are described in sections \ref{ssec:mm_image} and \ref{ssec:COline}. 

\subsection{Ancilliary data}

We compile a comprehensive set of photometric and spectroscopic data spanning optical to millimetre wavelengths. This includes optical and near-infrared photometry from \textit{Hipparcos} and 2MASS \citep{2000Hog,2006Skrutskie}, near- and mid-infrared photometry from \textit{Akari} and \textit{WISE} \citep{2010Ishihara,2010Wright},\, \textit{Spitzer} IRS spectroscopy and MIPS photometry \citep{2014Chen}, \textit{IRAS} and \textit{Herschel}/PACS far-infrared photometry \citep{2016Morales}, and the newly acquired ALMA millimetre photometry. Photometric measurements were colour corrected before modelling following the relevant instrument handbooks. A summary of the compiled photometry is presented in Table \ref{tab:photometry}. The \textit{Spitzer} IRS spectrum was taken from the CASSIS archive \citep{2011Lebouteiller} and scaled to the stellar photosphere model assuming a negligible infrared excess at 5~$\mu$m. Spectrophotometry using {\sc PyPhot} \citep{PyPhot} was carried out to check for consistency between the rescaled spectrum and the \textit{WISE} and MIPS photometric bands. The rescaled \textit{Spitzer} IRS spectrum is consistent with the photometry.

These data complement the non-detection in scattered light from VLT/SPHERE and the spatially resolved imaging observations in thermal emission at far-infrared (\textit{Herschel}/PACS) and millimetre (ALMA) wavelengths. They will be used to infer the stellar properties, in conjunction with the \textit{Gaia}-derived stellar parallactic distance, and thereafter modelled in conjunction with the images to determine the dust grain minimum size and size distribution exponent through radiative transfer modelling.

\section{Modelling and results}
\label{sec:mod}

Here we detail the data analysis and its outcomes. We first determine the fundamental parameters of HD~138965 to identify an appropriate stellar atmosphere model. 
The orientation and extent of the debris disc are then calculated from the ALMA imaging observations with constraints from the previous, lower angular resolution Herschel data. The stellar model in conjunction with the dust spatial distribution are used as inputs to a 1D radiative transfer code to infer the dust grain properties ($s_{\rm min}$, $M_{\rm dust}$, $q$) consistent with the observed thermal emission.

\subsection{Fundamental stellar parameters}
\label{ssec:star}
In order to derive the fundamental stellar parameters of HD\ 138965 (Gaia DR3 5819811291755502976; 2MASS J15401155-7013403; IRAS 15351-7003), we applied the Bayesian inference code of \citet{delburgo2016,delburgo2018} 
on a grid of stellar evolution models constructed from the PARSEC v1.2S tracks \citep{bressan2012,chen2014,chen2015,tang2014}. This choice is based on the good statistical agreement for detached eclipsing binaries, especially for stars on the main-sequence, where their measured dynamical masses and the corresponding predictions are consistent on average to within \,4\% \citep{delburgo2018}. 

The grid of PARSEC v1.2S models arranged for this analysis comprises ages ranging from 2 to 13,800 million years and steps of 5\%, and [M/H] from -2.18 to 0.51 with steps of 0.02 dex, adopting the photometric passband calibration of \citet{riello2021}, 
with the zero points of the VEGAMAG system. 
We used it to infer the stellar parameters of HD 138965 through the code of \citet{delburgo2016,delburgo2018}, fed by the three following input parameters: the absolute $G$ magnitude $M_{G}$, the colour $G_{\rm BP}-G_{\rm RP}$, and the iron to hydrogen [Fe/H]. Table \ref{tab:stellar_parameters} lists them. We reasonably assumed null extinction when deriving the colour $G_{\rm BP}-G_{\rm RP}$ and $M_{G}$ from the \textit{Gaia} DR3 photometry and astrometry \citep[][]{2023aGaia}. $M_{G}$ was determined from the apparent G magnitude by subtracting the distance modulus, which was calculated from the distance of 78.60$~\pm~$0.17 pc corresponding to the \textit{Gaia} DR3 trigonometric parallax. We adopted [Fe/H]=~$-0.30~\pm~0.30$ dex, 
derived by \citet{zhang2023} from \textit{Gaia} XP spectra. 

The Galactic space coordinates in units of parsec of HD~138965 are (X, Y, Z) = (55.66$~\pm~$0.11, -53.09$~\pm~$0.11, -16.195$~\pm~$0.032), which lies on the perimeter of the nearby, 40--50~Ma Argus association \citep[see][]{2019Zuckerman}. In addition, the Galactic space velocities in km s$^{-1}$ of HD~138965 are (U, V, W) = (-24.69$~\pm~$0.24, -10.58$~\pm~$0.23, -4.84$~\pm~$0.07), where U, V and W are positive toward the Galactic centre, the Galactic rotation, and the north Galactic pole, respectively. We determined the spatial and velocity parameters of HD~138965 from its \textit{Gaia} DR3 Equatorial Coordinates (ICRS) at Ep=2016.0, proper motion, radial velocity (-9.38$~\pm~$0.30 km s$^{-1}$) and parallax. \citet{2019Zuckerman} used \textit{Gaia} DR2 data instead, deriving a mean Galactic space velocity in km s$^{-1}$ of (-22.5$~\pm~$1.2, -14.5$~\pm~$2.1, -5.0$~\pm~$1.6), excluding HD~188728, CD-52 9381 and HD~192640 because they may not be members of Argus, as well as HD~129496, HD~145689, and even HD~138965 because of the relative large uncertainties in UVW, and adding HD~102647. We calculated a mean Galactic space velocity, in km s$^{-1}$, from \textit{Gaia} DR3 data without adding or removing any star, arriving at (-23.1$\pm$2.0, -14.5$~\pm~$2.7, -4.9$~\pm~$1.6). We note that \citet{2019Zuckerman} adopted a radial velocity of -2.0$~\pm~$4.3 km s$^{-1}$ for HD 138965, which is significantly lower and much less precise than the most recent value from \textit{Gaia} DR3. A full discussion of the membership of all stars collected by \citet{2019Zuckerman} is out of the scope of this paper, but we find convincing evidence in the \textit{Gaia} DR3-based spatial and velocity distributions that HD~138965 is particularly a member of the Argus association. This is supported by the Banyan Sigma analysis which assigns a 99.4\% probability of its membership \citep{2018Gagne}.
Imposing the prior that it is on the pre-main sequence and using the aforementioned inputs to feed the Bayesian inference code, we inferred the fundamental parameters of HD~138965 presented in Table \ref{tab:stellar_parameters}. 

Our 
effective temperature ($T_{\rm eff}$ = 8881$~\pm~$ 31\,K) and surface gravity ($\log\,g$= 4.332$~\pm~$0.012) agree within a 2-$\sigma$ level with those of \citet{zhang2023} ($T_{\rm eff}$ = 9814$~\pm~$442\,K, $\log\,g$= 4.69$~\pm~$0.20). 
Our age estimation of 31$~\pm~$11 Ma is consistent with that of \citet{2007Rhee} (20 Ma), who considered the UVW analysis of \citet{zuckerman2004} 
as well as the location of HD~138965 on the Hertzsprung-Russell diagram. 

As a validation test, we fed our Bayesian code with the same three input parameters but without imposing the prior on the evolution stage, finding that the age significantly increases to 805$~\pm~$468 Ma. This value is within 1-$\sigma$ uncertainty of the value of 350$^{+37}_{-56}$ Ma from \citet{2015Trevor}. The similarity is even greater if we adopt Solar metallicity, since we arrive at an age of 338$~\pm~$185 Ma. 
However, the rotational broadening $v \sin i$= 94.8$~\pm~$1.7 km s$^{-1}$ from \textit{Gaia} DR3 and the radius of HD~138965 lead to a maximum rotational period of 0.831$~\pm~$0.016 days that supports the youth of this host star.


\begin{table}
    \centering
    \caption{Fundamental stellar parameters of HD~138965.  \label{tab:stellar_parameters}}
    \begin{tabular}{lcc}
    \hline\hline 
        Parameter & Value & Ref. \\
    \hline
        $\Pi$ ($mas$) & 12.722 $\pm$ 0.028 & \textit{Gaia} DR3 \\   
        \hline
        $M_G$ (mag) & 1.949 $\pm$ 0.006 &  \textit{Gaia} DR3\\       
        $G_{BP}-G_{RP}$ (mag) & 0.087 $\pm$ 0.005 & \textit{Gaia} DR3 \\  
        {[Fe/H]}    & -0.3 $\pm$ 0.3 & \citet{zhang2023}\\

        \hline
        Effective temperature (K) & 8881 $\pm$ 31 & this work \\
        Radius ($R_{\odot}$) & 1.557 $\pm$ 0.009 & this work\\
        Mass ($M_{\odot}$) & 1.90 $\pm$ 0.04 & this work\\
        Surface gravity ($\log g$, cgs) & 4.332  $\pm$ 0.012 & this work\\
        Luminosity ($L$/$L_{\odot}$) & 13.59 $\pm$ 0.09 & this work\\
        Bolometric magnitude (mag) & 1.907 $\pm$ 0.007 & this work\\
        Age (Ma)   & 31 $\pm$ 11	& this work\\
    \hline
    \end{tabular}
\end{table}

\subsection{Millimetre continuum emission}
\label{ssec:mm_image}

We model the disc architecture as a single Gaussian belt. The model is defined by a flux density $f_{\rm disc}$, radius $R_{\rm peak}$, width (standard deviation) $\sigma_{R}$, scale height $h = z/R$, inclination $i$, and position angle $\phi$. Additionally, we include parameters for the stellar photospheric contribution $f_{\star}$, and position offsets $\Delta$RA, $\Delta$Dec between the star (assumed to be at the phase centre of the observations) and the disc centre. Previous marginally resolved \textit{Herschel} observations of the disc at far-infrared wavelengths provide good constraints on the radius, inclination, and position angle for the priors for these parameters \citep{2016Morales,2021Marshall}. These priors are normally distributed with a standard deviation 3\% of the initial value and bounds informed by the likely range of each parameter.

\begin{figure*}
	\includegraphics[width=\columnwidth,trim={0 2cm 0 0},clip]{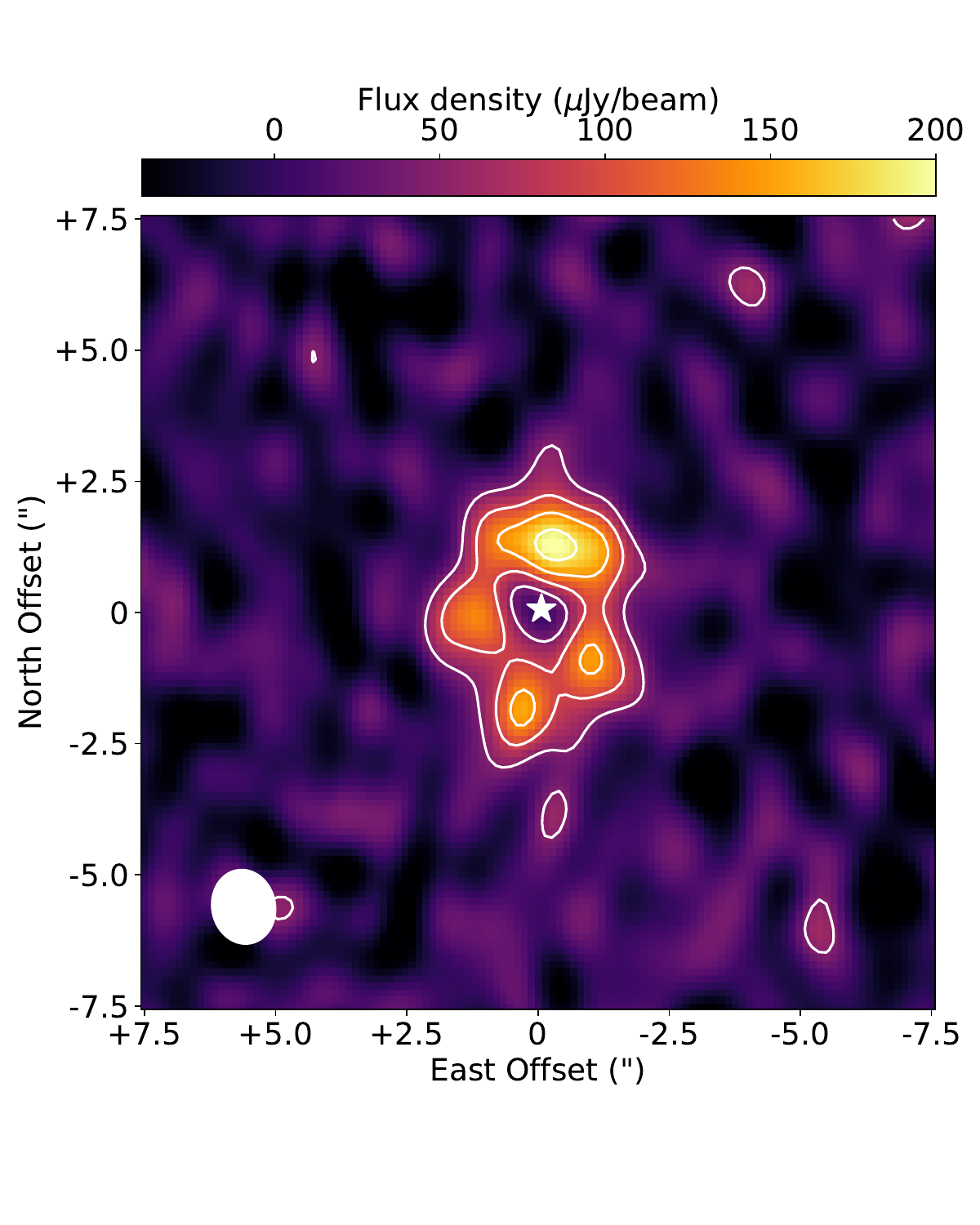}
	\includegraphics[width=\columnwidth,trim={0 2cm 0 0},clip]{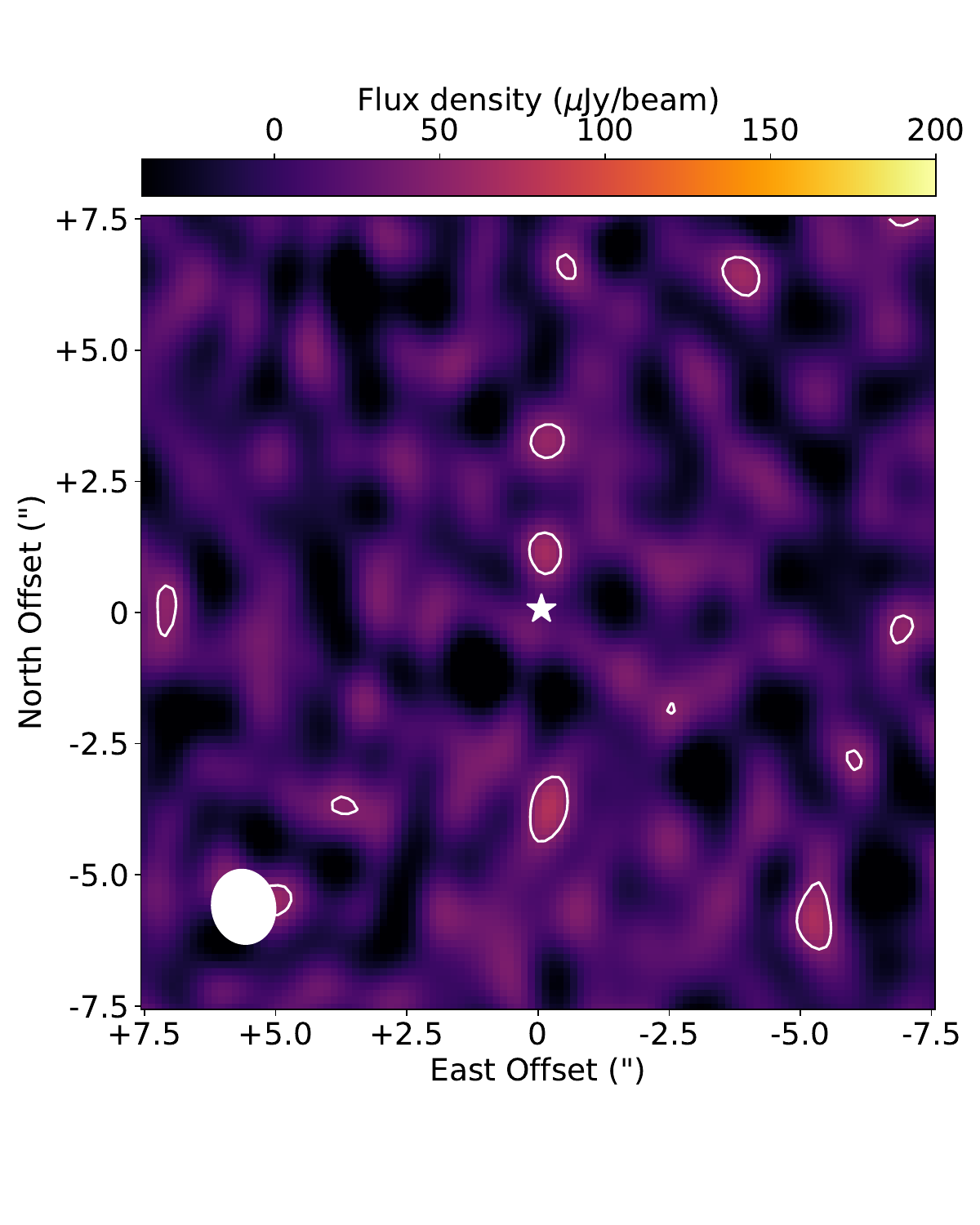}
    \caption{\textit{Left}: ALMA Band 6 continuum image of HD~138965. The image has been \textit{clean}ed and reconstructed with a Briggs weight of 0.5. The instrument beam ($1\farcs39~\times~1\farcs17$, $\phi = 10.5\degr$) is denoted by the white ellipse in the bottom left corner. Contours are in steps of 2-$\sigma$ from +2-$\sigma$. Orientation is north up, east left. \textit{Right}: Residuals after subtraction of the maximum likelihood model from the observations. We see there two features of interest in the residual map. A 2-$\sigma$ bump is located at the position of the northern ansa of the ring and is responsible for the observed `asymmetry', and another 2-$\sigma$ bump is located south of the ring exterior to the disc location. 
    }
    \label{fig:hd138965_alma}
\end{figure*}

We explore the parameter space of the model in a Bayesian framework using {\sc emcee} \citep{2013ForemanMackey}, following the approach presented in previous, similar works \citep[e.g.][]{2018Marshall,2023aMarshall}. Walkers are initialised at the maximum likelihood values for the disc parameters determined from modelling the \textit{Herschel} observations. Each set of parameters is used to generate a disc model using the radiative transfer code RADMC-3D \citep{2012Dullemond} which is then converted into synthetic visibilities using {\sc Galario} \citep{2018Tazzari}. These model visibilities are binned and compared to the ALMA observations through least-squares fitting. In total, we created 90,000 realisations of the model (90 walkers, 1,000 steps). The final posterior probability distribution used the last 10,000 realisations, from which the maximum likelihood parameters and their uncertainties were obtained. The posteriors are all monomodal and approximately normally distributed. We therefore take the 50th percentile of the posterior of each parameter as its model value and calculate the uncertainties from the 16th and 84th percentiles of each parameter distribution. We present the marginalised posteriors in Appendix \ref{AppendixB}. The results of the model fitting are summarised in Table \ref{tab:hd138965_alma_fit}, and we show the observations and residuals (after subtraction of the maximum likelihood model) in Figure \ref{fig:hd138965_alma}.

\begin{table}
	\centering
	\caption{Results from visibility fits to the ALMA observation. We note that the relative scale height $h = (z/R)$ is unconstrained from the modelling and the stellar flux \textbf{density} $f_{\star}$ is an upper limit.}
	\label{tab:hd138965_alma_fit}
	\begin{tabular}{lcc}
    	\hline\hline
		Parameter & Range & Value \\
    	\hline
    	$R_{\rm peak}$ (au) & 50 -- 300 & 150$^{+10}_{-7}$   \\
    	$\sigma_{R}$ (au)  & 10 -- 100 & 49$^{+7}_{-6}$   \\
    	$h$ & 0.01 -- 0.30 & 0.15$^{+0.07}_{-0.08}$ \\
    	$i$ (\degr) & 0 -- 90 & 49.9$^{+3.3}_{-3.7}$   \\
    	$\phi$ (\degr) & 0 -- 180 & 173.3$^{+3.7}_{-4.1}$   \\
    	$f_{\rm dust}$ (mJy) & 0.1 -- 10.0 & 1.460~$\pm$~0.230   \\
    	$f_{\rm star}$ ($\mu$Jy) & 0 -- 300 & $<~70$ \\
    	$\Delta$RA (\arcsec) & -1 -- 1 & 0.11$^{+0.5}_{-0.6}$ \\
        $\Delta$Dec (\arcsec) & -1 -- 1 & 0.16$^{+0.7}_{-0.7}$ \\
    	\hline
    \end{tabular}
\end{table}

We find that HD~138965's disc has a total flux density of 1.46~$\pm$~0.23~mJy at a wavelength of 1.269~mm. The cold belt is spatially resolved, with a radius of 150$^{+10}_{-7}$~au and a width of 49$^{+7}_{-6}$~au (FWHM = 115$^{+16}_{-14}$~au). The radius is consistent with the \textit{Herschel}-derived extent, and the width of the debris belt is marginally resolved with a fractional width $\Delta R / R = 0.77$. This places HD~138965 amongst the more radially extended discs that have been observed to-date spanning 0.1 to $>$1.0 \citep{2025Matra}, and in comparison to the Solar system's Edgeworth-Kuiper belt \citep[$\Delta R/R \simeq 0.3$][]{2009Kavelaars,2011Petit}. A 2-$\sigma$ brightness asymmetry between the NW and SE ansae of the disc is seen in the imaging data.  Given the large uncertainties and calibration of these observations it is not necessarily a real feature of the disc, but should motivate future observations. The large relative vertical scale height, $h = 0.15^{+0.07}_{-0.08}$, might be taken as a further indication of the disc being stirred. The associated uncertainty, in conjunction with the moderate inclination and modest signal-to-noise, instead leads us to the interpretation that a broad range of scale heights tested by the model will adequately fit the observations. This motivates the need for higher spatial resolution, higher signal-to-noise observations in future to constrain the internal dynamics of this disc. Its inclination, $i = 49\fdg9^{+3.3}_{-3.7}$, and position angle, $\phi = 173\fdg3^{+3.7}_{-4.1}$, are the most precise measurements of the disc orientation, both consistent with previous estimates. As expected we do not detect either emission from the warm belt or the star and obtain a 3-$\sigma$ upper limit of 90~$\mu$Jy on the stellar photospheric emission. Furthermore, we find no evidence for an offset between the star (phase centre) and disc centre from these observations although the constraints on such an offset are low due to the limited spatial resolution of the observations.

\subsection{Millimetre CO line emission}
\label{ssec:COline}

\begin{figure}
    \centering
    \includegraphics[width=\columnwidth]{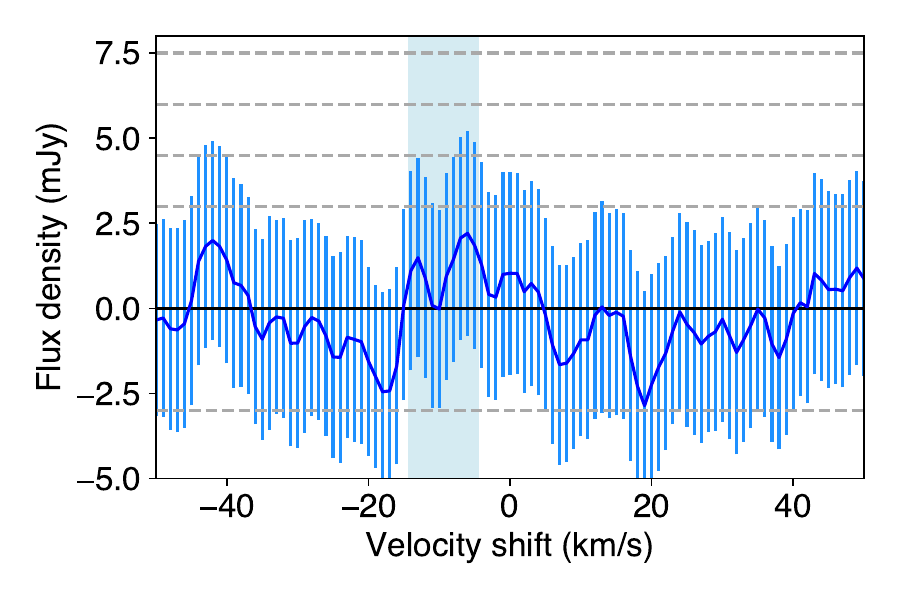}
    \caption{ALMA CO spectrum centred on a rest frequency of 230.538~GHz. The blue shaded region denotes the 10 km/s wide channel used to determine the presence, or rather absence, of significant CO emission, centred on the stellar radial velocity from \textit{Gaia} DR3 (i.e. -9.38 km/s). Uncertainties in the unbinned spectrum channels (1.3 km/s wide) are denoted by the blue vertical lines. The grey horizontal dashed lines denote the uncertainty in the unbinned spectrum from +2$\sigma$ in steps of 1-$\sigma$.}
    \label{fig:hd138965_alma_line}
\end{figure}

We use the spatially resolved extent and orientation of the disc from the continuum image to identify the region of the spectral window covering the CO (2-1) line at 230.538~GHz to search for gas emission from the disc. A spatial mask is applied to the spectral cube, with pixels lying within the region from $R_{\rm peak} -$HWHM to $R_{\rm peak} +$HWHM and with a flux density $\geq $3-$\sigma$ being included in the summation. For each pixel included in the mask we shift the spectrum's frequency according to the projected Keplerian velocity, interpolate these individual spectra to a set of common velocities and sum these to produce a final CO spectrum. The resulting spectrum centred on the CO (2-1) line at 230.54~GHz is presented in Figure \ref{fig:hd138965_alma_line}, with a shaded region denoting the frequency space included in the search for CO line emission centred at the stellar radial velocity of -9.38 km/s taken from \textit{Gaia} DR3 \citep{2023aGaia}. We find no evidence for any CO emission associated with the disc, and obtain a 5-$\sigma$ upper limit of 3.35 mJy/beam in a 10 km/s wide channel, 
equivalent to 2.7$\times10^{-23}$~W/m$^{2}$, which is around a factor of three above the predicted CO emission level of 1.1$\times10^{-23}$~W/m$^{2}$ ($M_{\rm CO}$ = 8.6$\times10^{-7}~M_{\oplus}$) of \citet{2017Kral}.

\subsection{Spectral energy distribution}
\label{ssec:sed}

We model HD\ 138965’s spectral energy distribution (SED) as a star plus two dust components, using two separate parametric models to represent the warm excess present at mid-infrared wavelengths, and the cold excess at mid-infrared to millimetre wavelengths. A summary of the photometry used in the radiative transfer modelling is provided in Table \ref{tab:photometry}.

\begin{table}
	\centering
	\caption{Photometry used in the radiative transfer modelling.}
	\label{tab:photometry}
	\begin{tabular}{cccc}
    	\hline\hline
		Wavelength & Flux Density & Telescope / &  Reference \\
		 ($\mu$m)  &    (mJy)     & Instrument / Filter &            \\
    	\hline
        0.44 & 10700~$\pm$~200 & Johnson $B$ & 1 \\
        0.55 & 9838~$\pm$~186  & Johnson $V$ & 1 \\
        0.58 & 8680~$\pm$~20   & \textit{Gaia} $G$ & 2 \\
        1.24 & 4654~$\pm$~172  & 2MASS $J$ & 3 \\
        1.65 & 3052~$\pm$~190  & 2MASS $H$ & 3 \\
        2.16 & 2059~$\pm$~58 & 2MASS $K_{\rm s}$ & 3 \\
        3.4  & 921~$\pm$~88  & \textit{WISE} W1 & 4 \\
        4.6  & 551~$\pm$~19  & \textit{WISE} W2 & 4 \\
        9    & 206~$\pm$~10  & \textit{Akari} IRC9 & 5 \\
        12   & 103~$\pm$~4   & \textit{WISE} W3 & 4 \\
        13   & 134~$\pm$~7   & \textit{Spitzer} IRS & 6 \\
        22   & 80~$\pm$~5    & \textit{WISE} W4 & 4 \\
        24   & 81~$\pm$~2    & \textit{Spitzer} MIPS & 6 \\
        31   & 111~$\pm$~2   & \textit{Spitzer} IRS & 6 \\
        60   & 482~$\pm$~63  & \textit{IRAS} 60 & 7 \\
        70   & 561~$\pm$~29  & \textit{Spitzer} MIPS &  6 \\
        100  & 527~$\pm$~17  & \textit{Herschel} PACS &  8 \\
        160  & 299~$\pm$~13  & \textit{Herschel} PACS &  8 \\
        1270 & 1.46~$\pm$~0.23  & ALMA Band 6 & 9 \\
    	\hline
    \end{tabular}
\raggedright
{\textbf{References:} 1. \cite{2000Hog} 2. \cite{2023aGaia} \\ 3. \cite{2006Skrutskie} 4. \cite{2010Wright} 5. \cite{2010Ishihara} \\ 6. \cite{2014Chen} 7. \cite{1988Beichman} 8. \cite{2016Morales} \\ 9. This work.}
\end{table}

We obtained the best BT-NextGen stellar atmosphere model \citep[AGSS2009,][]{2012Allard} that fits the HD~138965's photometry from a classical least-squares approach. The fundamental stellar parameters derived are the effective temperature $T_{\rm eff} = 8800$~K, metallicity [Fe/H] = -0.5, and the surface gravity of $\log g = 4.5$, consistent with the values inferred in Section \ref{ssec:star}. This model was scaled to the available optical and near-infrared photometric data up to 5~$\mu$m.

To model the thermal emission of the grains we applied \textit{Stardust}, a model that takes into account the dust composition (determined by optical constants) and a combination of forces that influence the spatial redistribution of dust grains. These forces include the conservative forces of gravity competing with the outward radiation pressure \citep[e.g.,][]{1979BurnsLamySoter} , and the non-conservative force of the Poynting-Robertson effect \citep[e.g.,][]{1903Poynting,1937Robertson,2007Klacka} that causes grains to spiral inwards. \textit{Stardust} includes modelling a simple collisional lifetime of the grains that depends on the optical depth of the debris disc \citep[e.g.,][]{2010KuchnerStark}. We used the fractional luminosity of HD 138965’s disc as a proxy for the optical depth. We summarise the \textit{Stardust} model in Appendix \ref{AppendixA}.

For our purposes, we assumed the radial planetesimal surface density to follow a Gaussian profile and differential power law grain size distribution ($dn/ds \propto s^{-q}$). In all considered scenarios, we assume the maximum size of the dust grains to be 3000 $\micron$ and the inner belt to have a grain composition of astronomical silicate \citep{2003Draine}. The outer, cooler belt was assumed to have a fixed peak surface density at 150~au with a standard deviation of 49~au (taken directly from the ALMA observations, see Table \ref{tab:hd138965_alma_fit}).

We note that, since the inner belt is not spatially resolved, there will be a degeneracy between location and the temperature of the grains.  We performed an initial analysis using \textit{Stardust} to estimate the  inner belt characteristics.  As such we had the following components to be free parameters for both the inner and outer belts: belt radius ($r_m$) and width ($\sigma_r$), minimum grain size ($s_{\rm min}$), dust mass ($M_s$), and the size distribution component ($q$) to initially vary between 3 and 4 as is the case for collisional active discs \citep[e.g.,][]{2016Macgregor,2017Marshall,2021Norfolk}.  Here we found that the inner belt had a minimum grain size of approximately 12 $\micron$, however, as the value of $q$ did not converge.

To pin down the radial component of the inner belt, we ran \textit{Stardust} with the following assumptions for the inner belt: minimum grain size of 12 $\micron$, size distribution component to be 3.5 \citep[steady-state collisional cascade;][]{1969Dohnanyi}, and the disc fractional width ($\Delta r/r$) to be 0.3.  The mean radial distance of the inner belt and the mass were kept as free parameters.  This scenario still did not converge to a solution, but it did produce a reasonable goodness-of-fit to the data where our reduced minimum $\chi^2$ test was less than one. The posterior probability distribution for the corresponding {\sc emcee} run is presented in Figure \ref{fig:hd138965_SED_corner_plot} (Appendix \ref{AppendixB}). A summary of the fixed parameters that have been assumed for the inner belt component of HD 138965 are given in Table \ref{tab:hd138965_innerbelt}.

\begin{table}
	\centering
	\caption{Assumed inner belt parameters for HD 138965.}
	\label{tab:hd138965_innerbelt}
	\begin{tabular}{lc}
    	\hline\hline
		Parameter & Value \\
    	\hline
        $s_{\rm min}$ ($\micron$) & 12 \\ 
        $T_{dust}$ (K) & 168 \\
        $q$ & 3.5 \\
        $M_{\rm s}$ ($\times10^{-3}~M_{\oplus}$) & 0.154 \\
    	$r_m$ (au) &   13.7 \\
        $\sigma_{r}$ (au)  & 1.75  \\ 
        Composition & Astronomical Silicate\\
     	\hline
    \end{tabular}
\end{table}

Now that we have our fixed assumptions for the inner belt, we then modelled the outer belt with different compositions and porosities.  For all scenarios, we set the minimum grain size in the outer belt to be the blowout limit of the corresponding grain composition.  For the outer belt component ($q$), we allowed to vary between 2 and 5 to account for potential size-dependent velocity distributions for all bodies \citep[e.g.,][]{2012PanSchlichting}.  The different compositions considered were a mixture of astronomical silicate with either crystalline water-ice \citep{2008WarrenBrandt}, amorphous water-ice \citep{1993Hudgins}, amorphous carbon \citep{1993Preibisch}, or vacuum (as porosity).  For each mixture (or inclusion matrix particles; IMPs), we assumed the grains to be astronomical silicate combined with an inclusion material with a volume fraction of 0.5, 0.25, and 0.1 for each material.

To determine the maximum likelihood values for each model we again adopt a Bayesian approach using the Markov Chain Monte Carlo (MCMC) code {\sc emcee} \citep{2013ForemanMackey} to explore the parameter space and determine the posterior probability distributions for each parameter from which the maximum amplitude probability and uncertainties were determined.  We ran a total of 320\,000 realisations for each model with 80 walkers and 4\,000 steps. We used the first 500 steps as burn-in for the MCMC chains and calculated probability distributions from the final 280\,000 realisations.  We used the 50$^{\rm th}$ value from the posterior of each parameter to determine the reduced $\chi^2$ of a given architecture and dust composition for comparison between dust composition models. For each of these scenarios, all of them converged to a solution before reaching the 4000th step.

To determine which model to be the best fit, we applied the Bayesian Information Criterion \citep[BIC,][]{1978Schwarz}:

\begin{equation} \label{eq:BIC}
    BIC  = \chi^2 + k \ln N,
\end{equation}

where `$k$' is the number of free parameters and `$N$' is the number of data points used in the fit.  In our scenarios, we have $k = 3$ (for outer belt composed of only astronomical silicate) or $k = 4$ (for outer belt composed of mixed grains), and $N = 15$.  The lower the BIC value the better the fit.  We then calculate the change of BIC \citep[$\Delta BIC$,][]{2004Burnham}:

\begin{equation} \label{eq:dBIC}
    \Delta BIC  = BIC_{mc} - BIC_{best},
\end{equation}

where `$mc$' is the model being compared with the `$best$' model.  In our case, the best model as determined by `BIC' is the scenario where the outer belt is composed of pure astronomical silicate dust grains.  If $\Delta BIC < 2$, the models are indistinguishable; $2 < \Delta BIC < 6$ there is some support for the `best' model but not conclusive; $6 < \Delta BIC < 10$ there is stronger evidence to support the best model; ands $\Delta BIC > 10$ indicates that the `best' model is highly preferred. See Table \ref{tab:au-beta} for a summary of the model outcomes.

\begin{figure}
    \centering
    \includegraphics[width=0.5\textwidth]{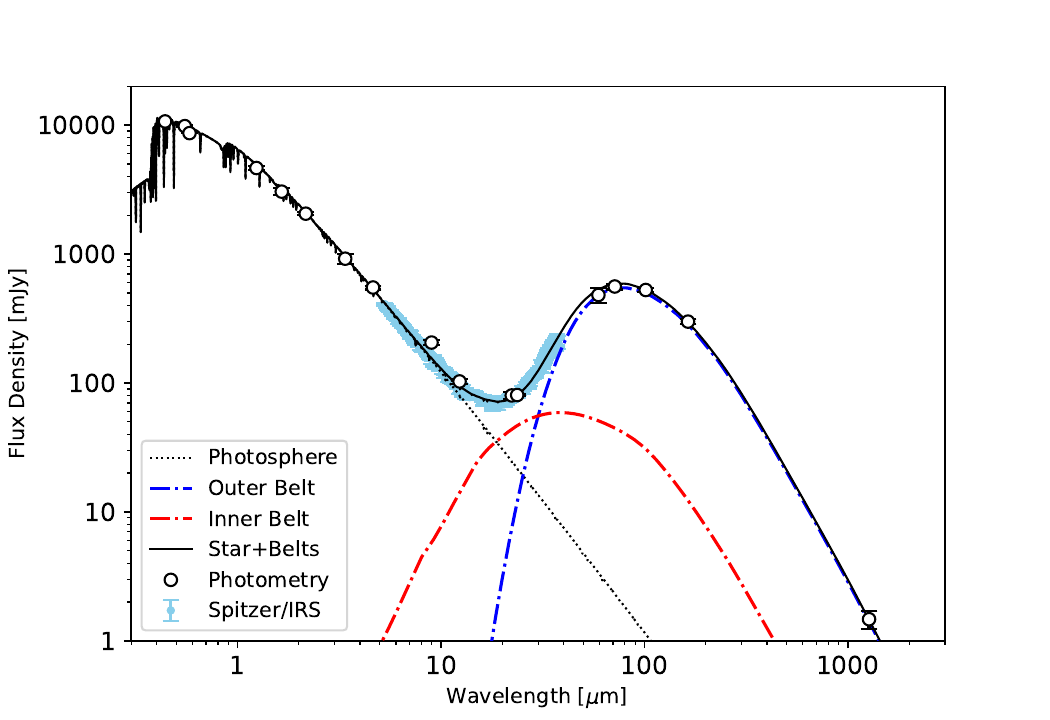}
    \caption{Spectral energy distribution of HD~138965 for the scenario where grains subjected to forces of gravity, radiation pressure, and drag induced by the Poynting-Robertson effect. The inner and outer belts were composed of 90:10 astronomical silicate:amorphous water-ice. White data points are literature photometry, except for the ALMA data point. The light blue line denotes the \textit{Spitzer}/IRS spectrum. The stellar photosphere model is denoted by the black dotted line, and the total emission by the black solid line. The blue and red dashed-dot lines denote the cold and warm disc components, respectively.}
    \label{fig:sed}
\end{figure}

On this basis, the model corresponding to P = 0.25 is a reasonable fit to the observations, while 90:10 astronomical silicate:amorphous water-ice, and the remaining porosity scenarios are all plausible dust compositions. See Figure \ref{fig:sed} for the corresponding SED to the scenario of 90:10 astronomical silicate:amorphous water-ice. The 90:10 and 75:25 of astronomical silicate:crystalline water-ice were not categorically ruled out based on the $\Delta$BIC values. We can however rule out the 50:50 compositions for both amorphous and crystalline water ice, and all three of the carbon:astronomical silicate compositions tested here. It should be noted that the emission from the inner belt is fixed following the initial modelling where the outer belt grains are assumed to be solely astronomical silicate. This may introduce an inherent bias toward astronomical silicate being the best solution for the outer belt. Ultimately, we would need to either independently determine the architecture of the inner belt through spatially resolved imaging or obtain spectroscopic data for the dust revealing features associated with water ice or other species to better infer the composition of the debris dust.

Through our initial analysis using MCMC, we found the value of the inner belt's mean radial location to be at $15^{+3}_{-2}$ au, and minimum grain size $s_{\rm min}$ of 12$\micron$, broadly consistent with the values reported in \cite{2016Morales}. 
The $s_{\rm min}$ adopted for the inner belt is much larger than that of the outer belt. We would expect $s_{\rm min}$ to be a few times the blowout size and for a 13.6~$L_{\odot}$ star, that ratio would be $s_{\rm min} \simeq 2.5 s_{\rm blow}$ or 7~$\mu$m \citep{2014Pawellek}.
The effective temperature of 12~$\mu$m astronomical silicate dust grains is 168~K at 15~au from HD~138965, consistent with the sublimation temperature of water ice in vacuum \citep[e.g.][]{2021Kossacki}. This supports our assumption of a dry inner belt; if the minimum grain size were smaller the dust would be hotter such that the major implication would be the inner belt would need to be placed further out to match the observed mid-infrared emission.
Our study of the outer belt composition implemented a refined model compared to \cite{2016Morales}, where they assumed a single composition of 50:50 astronomical silicate:ice for the dust grains and a fixed density distribution. Our best-fit results correspond to the porosity scenarios and have a strong candidate with 90:10 astronomical silicate:crystalline water-ice mixture for the outer belt.

\begin{table*}
\setlength{\tabcolsep}{2pt} 
	\caption{Input parameters and the corresponding goodness-of-fit values from the radiative transfer modelling. S: astronomical silicate; WC: crystalline water ice; WA: amorphous water ice; C: carbon; P: porosity; with the volume fraction of inclusion material mixed with the astronomical silicate are 0.10, 0.25 and 0.50 respectively. NB: Dust masses are for grains up to 3~mm in size.}
	\label{tab:hd138965_sed_fit}
    \centering
    \renewcommand{\arraystretch}{1.5}
    \begin{tabular}{cccccccccccccc}
    \hline
    Parameter                     & S100                  & S90WC10              & S75WC25              & S50WC50              & S90WA10              & S75WA25              & S50WA50              & S90C10                & S75C25               &S50C50                & P = 0.1                &  P = 0.25         & P = 0.5         \\
    \hline
    $s_{\rm blow}$ ($\micron$)        & 2.70                   & 2.86                   & 3.16                   & 4.04                   & 2.86                   & 3.16                   & 3.87                   & 2.86                    & 3.04                   & 3.46                   & 2.93                   & 3.41                   & 4.80                  \vspace{0.2cm}\\
    $q$ [2.0 - 5.0]               & 3.73$^{+0.07}_{-0.07}$ & 3.82$^{+0.09}_{-0.08}$ & 3.86$^{+0.11}_{-0.10}$ & 4.01$^{+0.20}_{-0.17}$ & 3.78$^{+0.08}_{-0.08}$ & 3.83$^{+0.10}_{-0.09}$ & 4.00$^{+0.18}_{-0.15}$ & 4.15$^{+0.10}_{-0.09}$  & 4.41$^{+0.10}_{-0.09}$ & 4.89$^{+0.08}_{-0.16}$ & 3.63$^{+0.07}_{-0.07}$ & 3.49$^{+0.07}_{-0.06}$ & 3.30$^{+0.07}_{-0.07}$\vspace{0.2cm}\\
    $M_{\rm s}$ [1 - 20]          & 7.74$^{+0.49}_{-0.48}$ & 7.75$^{+0.55}_{-0.54}$ & 7.18$^{+0.58}_{-0.56}$ & 5.86$^{+0.70}_{-0.62}$ & 7.44$^{+0.51}_{-0.50}$ & 7.07$^{+0.55}_{-0.54}$ & 6.12$^{+0.65}_{-0.61}$ & 7.05$^{+0.43}_{-0.42}$  & 9.20$^{+0.76}_{-0.74}$ & 8.89$^{+0.60}_{-0.63}$ & 6.67$^{+0.43}_{-0.43}$ & 5.23$^{+0.35}_{-0.34}$ & 3.18$^{+0.21}_{-0.21}$\vspace{-0.1cm}\\
    ($\times10^{-3}~M_{\oplus}$)  &                        &                        &                        &                        &                        &                        &                        &                         &                        &                        &                        &                        &                       \\
    $\chi^2$         & 9.3                    & 14                   & 17                   & 150                  & 11                   & 28                   & 100                  & 19                    & 20                   & 120                  & 9.4                    & 8.3                    & 10                  \vspace{0.2cm}\\
    $\chi^2_{\rm red}$ & 0.66                   & 0.90                 & 1.1                  & 9.8                  & 0.78                 & 1.9                  & 6.9                  & 1.3                   & 1.3                  & 7.7                  & 0.62                   & 0.55                   & 0.67                   \vspace{0.2cm}\\ 
    $\Delta$BIC       & 0.0                    & 7.0                  & 10                   & 140                  & 4.4                  & 21                   & 96                   & 12                    & 14                   & 110                  & 2.8                    & 1.7                    & 3.4                   \vspace{0.2cm}\\ 
    \hline
    \label{tab:au-beta}
    \end{tabular}
\end{table*}

\subsection{Constraints from scattered light}
\label{ssec:scat_limits}

The large extent and broad width of HD~138965's debris disc combined with its intermediate inclination ($i = 49\fdg$9) result in a low surface brightness for the disc, despite its relatively high fractional luminosity ($L_{\rm d}/L_{\star} \simeq 4\times10^{-4}$). This makes it a challenging target for detection in scattered light. Previous attempts have been made with \textit{HST}/ACS at optical wavelengths and with VLT/SPHERE in the near-infrared. These have both failed to detect any emission from the debris disc. Using equation 2 from \cite{2018Marshall} in conjunction with the resolved disc architecture and known stellar brightness we can calculate a limit on the dust albedo, $\omega$. A point of caution must be made here in that the fractional luminosity ($L_{d}/L_{\star}$) which is used to infer the albedo in \cite{2018Marshall} may be a poor approximation of the absorption efficiency at the observing wavelength of scattered light imaging. The dust albedo calculated in that work is a bulk measure across all dust grain sizes, not the albedo for a specific wavelength or grain size.

HD~138965 was observed by VLT/SPHERE IRDIS \citep{2019Beuzit} using the BB\_H filter for the SPHERE High-Angular Resolution Debris Disks Survey\footnote{ESO programs 096.C-0388 and 097.C-0394} \citep[SHARDDS,][]{2016Wahhaj,2017Choquet,2018Marshall,2021Cronin-Coltsmann}. These observations are deeper than those presented in \cite{2018Matthews}, but the disc is still not detected. From the SHARDDS observation we obtain a 5-$\sigma$ upper limit of 83~$\mu$Jy/arcsec$^{2}$ at a separation of $2\arcsec$ from the star. To estimate the equivalent integrated disc brightness required to be detectable, we inject model discs with an architecture consistent with the ALMA observation into the data cube. The cube was then reduced and stacked using the classical ADI algorithm. The model brightness was varied until the integrated stack exhibited residuals in excess of the 5-$\sigma$ limit. We find a minimum brightness of around 10~mJy is required for the disc to be detected in the SHARDDS data. Based on this calculation, we would need to go at least a factor of four deeper in imaging HD~138965 to detect the disc at near-infrared wavelengths. This assumes reducing the limit of the albedo to $\leq 0.2$ would start to probe values consistent with other discs at near-infrared wavelengths \citep[e.g. ][]{2023Ren}.

\begin{figure}
    \centering
    \includegraphics[width=0.5\textwidth]{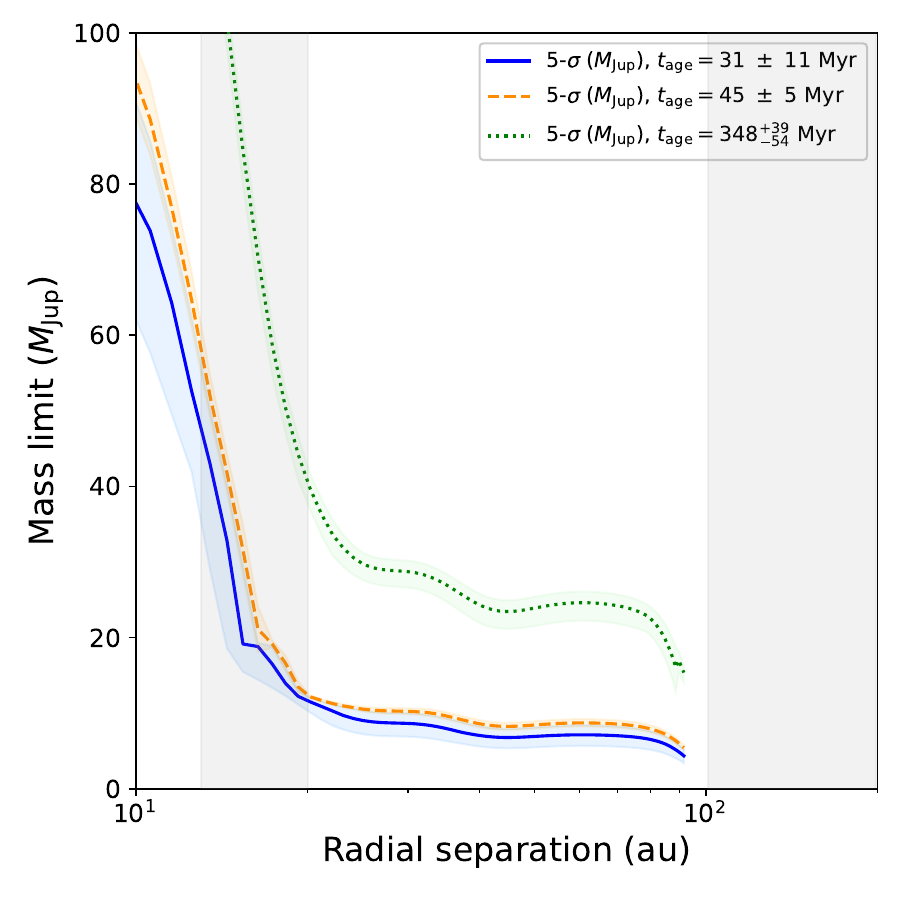}
    \caption{Mass limits from VLT/SPHERE SHARDDS observation of HD~138965 based on AMES-COND evolutionary models \citep{2001Allard}. Contrast as a function of separation was calculated using the RSM algorithm as presented in \citet{2022Dahlqvist}. The dot (green), dash (orange), and solid (blue) lines represent different assumed ages for the star, with the coloured shaded regions being the associated uncertainties. The grey shaded regions denote the expected extent of the two debris belts. We find limits at 80~au of 4~$M_{\rm Jup}$ for an age of 30~Ma, or 5.5~$M_{\rm Jup}$ for an age of 45~Ma.}
    \label{fig:contrast}
\end{figure}

We further estimate the contrast for point sources around HD~138965 at radial separations out to $\simeq~1\arcsec$. This contrast curve was calculated using the Regime Switching Model algorithm as presented in \cite{2022Dahlqvist}. We convert the observed contrast to mass in order to determine the limits on planetary companions in the vicinity of the disc, as shown in Figure \ref{fig:contrast}. The mass limit as a function of separation was calculated for ages of 30~$\pm$~10~Ma (our inferred age), 45~$\pm$~5~Ma (Argus association membership), and 348$^{+39}_{-54}$~Ma (from \textit{Gaia}). The masses were converted from the measured contrast using the AMES-COND grid of models \citep{2001Allard}. In the vicinity of the inner edge of the cold debris belt ($\simeq~80~$au) we obtain upper limits to the mass of any companion of 4$~M_{\rm Jup}$ ($t_{\star} = $~31~Ma), 5$~M_{\rm Jup}$ ($t_{\star} = $~45~Ma), and 15$~M_{\rm Jup}$ ($t_{\star} = $~348~Ma).

A surface brightness limit of 20~mag/arcsec$^{2}$ at $2\arcsec$ ($\equiv 33~\mu$Jy/arcsec$^{2}$) was obtained in the \textit{HST}/ACS F606W filter \citep{2010Krist}. The \textit{HST}/ACS measurement is more constraining than the VLT/SPHERE measurement, resulting in a limit of $\omega \leq 0.09$ at optical wavelengths. This is consistent with the minimum grain size in the outer belt derived from modelling the SED, assuming an astronomical silicate composition for the dust. The calculated albedo limit for HD~138965's disc is intermediate for those discs detected in scattered light \citep[0.02 to $\geq$0.2,][]{2018Choquet,2023Ren}, from which we infer deeper observations would stand a good chance of detecting the disc.

We can obtain an alternative measure of the dust scattering albedo using the maximum likelihood dust model determined in Section \ref{ssec:sed}. Here we take the $Q_{\rm abs}$ and $Q_{\rm sca}$ values derived from the dirty ice dust composition for grains with $s = 3.6~\mu$m, being the minimum grain size in our distribution and therefore the largest contribution to the disc surface area. We infer a scattering albedo of $\omega = 0.56$ over all angles at a wavelength of 1.6~$\mu$m, using the relation $\omega = Q_{\rm sca}/(Q_{\rm abs} + Q_{\rm sca})$ \citep{1998BohrenHuffman}. Considering instead only a restricted range of scattering angles close to $90\degr$, based on the orientation of the disc, we instead infer an effective dust albedo $\simeq 0.1$ for astronomical silicate dust grains around $10~\mu$m in size \citep[following e.g.][]{2013Mulders,2018Choquet}. This would result in non-detection of the disc and be marginally consistent with upper limits from \textit{HST}.

\section{Discussion}
\label{sec:dis}

The outer belt of HD~138965's debris disc is found to be very broad with $\Delta R / R = 0.77$. This is far greater than the Solar system's Edgeworth-Kuiper belt \citep[$\Delta R / R \simeq 0.3$,][]{2009Kavelaars,2011Petit} and comparable to HR~8799's outer belt \citep{2021Faramaz}. 
A handful of radially broad discs followed-up at sufficient angular resolution have been found to contain substructures associated with disc-planet interaction believed to represent sculpting or scattering of planetesimals \citep[e.g.][]{2021Marino}. The pursuit of further observations of HD~138965 at higher angular resolution to determine the architecture of the outer belt are therefore recommended.

Despite the modest angular resolution and signal-to-noise of the ALMA observations the outer belt is spatially resolved in both radius and width. We can therefore place constraints on the presence of a planetary companion stirring or sculpting the belt following the analysis presented in \cite{2022Pearce}. For these calculations we assume the disc inner edge to lie at 101~au (i.e. $R$ - $\sigma_{\rm R}$) and the stellar parameters from Table \ref{tab:stellar_parameters}. For the single planet sculpting scenario, we calculate a planetary mass $\geq$~2.3~$\pm$~0.4~$M_{\rm J}$ at a semi-major axis $\leq~74^{+8}_{-6}$~au. This is a factor of five improvement over the previous limit from \cite{2018Matthews}, and a factor of two better than obtained from the deeper SHARDDS image presented in Section \ref{ssec:scat_limits}, but still relatively weak. At the other extreme, assuming the disc is self-stirred by massive planetesimals, we can constrain the total mass of the debris disc to be at least $400~\pm~100~M_{\oplus}$ in bodies up to 100~km in size. This mass is consistent with the maximum allowable debris disc masses as a function of disc properties (e.g. radius, fractional luminosity) determined by \cite{2021KrivovWyatt}.

Our analysis of the dust dynamics has been modelled by \textit{Stardust}, where we include P-R effected grains (i.e., inward spiralling of dust grains) to eventually result in collisional destruction on its way to the host star. Due to the inner belt not being spatially resolved we attempted to mitigate the degeneracy between grain temperature and location, we performed an initial analysis to estimate its characteristics under the assumption that the outer belt grains are composed of pure astronomical silicate. The mean values taken from the posterior probability distributions for this scenario, we estimated the location of the inner belt to be mean distance of $15^{+3}_{-2}$~au, this result is in line with the study conducted by \cite{2016Morales}.  With all subsequent scenarios modelled, we fixed the minimum grain size of the outer belt to be the blowout limit of the grain corresponding to the relative dust composition. We note that under the assumptions of inner belt characteristics there may be an inherent bias towards the best model where pure astronomical silicate grains are in the outer belt. On this basis, we cannot rule out the scenarios where there is at least 10\% crystalline or amorphous water-ice inclusions in the astronomical silicate grains in the outer belt. However, we can eliminate grains for the scenarios with 50\% water-ice inclusions and any grain that has amorphous carbon mixed in for {HD~138965's} outer belt. To be sure of the inner belt radial location, higher spatial resolution images would be required. For example, at a radius of $\simeq 20$~au the inner belt would have an angular separation of $0\farcs25$ which is potentially resolvable by JWST/MIRI in imaging mode at 10~$\mu$m \citep{2023Wright}, or in future with AtLAST at 350~$\mu$m \citep{2024Klaassen,2024Booth}.


\section{Conclusions}
\label{sec:con}

We have spatially resolved for the first time the radius and width of the outer cool belt of HD 138965's debris disc at millimetre wavelengths with ALMA. We find the peak emission from the belt at 150~au with a standard deviation of 49~au (FWHM 115 au, $\Delta R /R \simeq 0.77$). The disc radius, inclination (49$\fdg$9), and position angle (173$\fdg$3) of the belt are in good agreement with previous marginally resolved \textit{Herschel} observations. The disc is very broad, comparable in width to several systems that have exhibited indirect evidence of disc-planet interactions. However, the low signal-to-noise and relatively low angular resolution of the ALMA observations preclude any more detailed insights regarding the belt structure or its stirring.

Using the newly acquired ALMA imaging of the disc we modelled the disc spectral energy distribution using a novel code to redistribute dust grains based on their size and the gravitational and radiation forces acting upon them. As the inner belt is not spatially resolved, we assumed that both the inner and outer belt grains were composed of pure astronomical silicate to constrain the system's inner belt by determining its location to be 15$^{+3}_{-2}$~au to help mitigate the degeneracy between the temperature and location for grains. We found, whilst the pure astronomical silicate grains were considered the best scenario for the composition of the dust in the outer belt, we could not categorically disregard scenarios with a 10\% crystalline or amorphous water-ice inclusions in the silicate grains.

We then undertook a study of archival scattered light observations from VLT/SPHERE to obtain some constraint on the dust optical properties, but the disc remains undetected. Using the non-detection by \textit{HST} at optical wavelengths in conjunction with the spatially resolved architecture of the disc from ALMA we calculate an upper limit to the dust albedo of 0.09. That limit from observations is inconsistent with the same limit determined from the dust grain size and composition inferred from the SED. These two values can be somewhat reconciled by considering either a restricted range of scattering angles to reduce the inferred albedo from HST imaging, or instead invoking the broad belt width and self-subtraction inherent to the data reduction process for its non-detection in SPHERE imaging.

Despite non-detection of the disc in scattered light, analysis of these data provided deeper constraints on the mass of companions to the system. Interpretation of the contrast curve derived from the SHARDDS observation of HD~138965 places a mass limit of 4~$M_{\rm Jup}$ at 80~au from the star, assuming an age of 30~Ma. Using the architecture of the disc we instead obtain a mass limit of 2.3~$\pm$~0.4~$M_{\rm Jup}$ for a single planet sculpting the disc. This comparison illustrates the predictive power of debris disc architectures for the inference of planetary companions.

Future avenues for the characterisation of the HD~138965 system with current and planned facilities are promising. For example, in the infrared \textit{JWST} is able to resolve both the inner and outer belts of the debris disc and simultaneously provide deep limits on the presence of planetary companions. At millimetre wavelengths, deeper observation with ALMA to resolve the outer belt's structure is possible, albeit demanding, in terms of the time required with its current sensitivity; future bandwidth upgrades will make such high angular resolution observations more feasible.

\section*{Acknowledgements}

This research has made use of the SIMBAD database, operated at CDS, Strasbourg, France \citep{2000Wenger}. This research has also made use of NASA's Astrophysics Data System.
This paper makes use of the following ALMA data: ADS/JAO.ALMA\#2019.1.01220.S. ALMA is a partnership of ESO (representing its member states), NSF (USA) and NINS (Japan), together with NRC (Canada), NSTC and ASIAA (Taiwan), and KASI (Republic of Korea), in cooperation with the Republic of Chile. The Joint ALMA Observatory is operated by ESO, AUI/NRAO and NAOJ.

JPM acknowledges support by the National Science and Technology Council of Taiwan under grant NSTC 112-2112-M-001-032-MY3.

CdB acknowledges support from the Agencia Estatal de Investigación del Ministerio de Ciencia, Innovación y Universidades (MCIU/AEI) under grant WEAVE: EXPLORING THE COSMIC ORIGINAL SYMPHONY, FROM STARS TO GALAXY CLUSTERS and the European Regional Development Fund (ERDF) with reference PID2023-153342NB-I00/10.13039/501100011033, as well as from a Beatriz Galindo Senior Fellowship (BG22/00166) from the MICIU. The Universidad de La Laguna (ULL) and the Consejería de Economía, Conocimiento y Empleo of the Gobierno de Canarias are also gratefully acknowledged for the support provided to CdB (2024/347).

This work was also partly supported by the Spanish program Unidad de Excelencia María de Maeztu CEX2020-001058-M, financed by MCIN/AEI/10.13039/501100011033.

EC acknowledges funding from the European Union (ERC, ESCAPE, project No 101044152). Views and opinions expressed are, however, those of the author(s) only and do not necessarily reflect those of the European Union or the European Research Council Executive Agency. Neither the European Union nor the granting authority can be held responsible for them.

VFG acknowledges funding from the National Aeronautics and Space Administration through the Exoplanet Research Program under Grants No. 80NSSC21K0394 and 80NSSC23K1473 (PI: S. Ertel), and Grant No 80NSSC23K0288 (PI: V. Faramaz).

\textit{Facilities}: ALMA

\textit{Software}: AstroPy \citep{2013AstroPy,2018AstroPy,2022AstroPy}, Corner \citep{2016ForemanMackey}, Emcee \citep{2013ForemanMackey}, Galario \citep{2018Tazzari}, Matplotlib \citep{2007Hunter}, Numpy \citep{2020Harris}, Scipy \citep{2020Virtanen}

\section*{Data Availability}

The data presented in this work are available upon reasonable request to the authors.



\bibliographystyle{mnras}
\bibliography{hd138965_refs} 

\begin{thebibliography}{}
\makeatletter
\relax
\def\mn@urlcharsother{\let\do\@makeother \do\$\do\&\do\#\do\^\do\_\do\%\do\~}
\def\mn@doi{\begingroup\mn@urlcharsother \@ifnextchar [ {\mn@doi@} {\mn@doi@[]}}
\def\mn@doi@[#1]#2{\def\@tempa{#1}\ifx\@tempa\@empty \href {http://dx.doi.org/#2} {doi:#2}\else \href {http://dx.doi.org/#2} {#1}\fi \endgroup}
\def\mn@eprint#1#2{\mn@eprint@#1:#2::\@nil}
\def\mn@eprint@arXiv#1{\href {http://arxiv.org/abs/#1} {{\tt arXiv:#1}}}
\def\mn@eprint@dblp#1{\href {http://dblp.uni-trier.de/rec/bibtex/#1.xml} {dblp:#1}}
\def\mn@eprint@#1:#2:#3:#4\@nil{\def\@tempa {#1}\def\@tempb {#2}\def\@tempc {#3}\ifx \@tempc \@empty \let \@tempc \@tempb \let \@tempb \@tempa \fi \ifx \@tempb \@empty \def\@tempb {arXiv}\fi \@ifundefined {mn@eprint@\@tempb}{\@tempb:\@tempc}{\expandafter \expandafter \csname mn@eprint@\@tempb\endcsname \expandafter{\@tempc}}}

\bibitem[\protect\citeauthoryear{{Allard}, {Hauschildt}, {Alexander}, {Tamanai}  \& {Schweitzer}}{{Allard} et~al.}{2001}]{2001Allard}
{Allard} F.,  {Hauschildt} P.~H.,  {Alexander} D.~R.,  {Tamanai} A.,   {Schweitzer} A.,  2001, \mn@doi [\apj] {10.1086/321547}, \href {https://ui.adsabs.harvard.edu/abs/2001ApJ...556..357A} {556, 357}

\bibitem[\protect\citeauthoryear{{Allard}, {Homeier}  \& {Freytag}}{{Allard} et~al.}{2012}]{2012Allard}
{Allard} F.,  {Homeier} D.,   {Freytag} B.,  2012, \mn@doi [Philosophical Transactions of the Royal Society of London Series A] {10.1098/rsta.2011.0269}, \href {https://ui.adsabs.harvard.edu/abs/2012RSPTA.370.2765A} {370, 2765}

\bibitem[\protect\citeauthoryear{{Astropy Collaboration} et~al.,}{{Astropy Collaboration} et~al.}{2013}]{2013AstroPy}
{Astropy Collaboration} et~al., 2013, \mn@doi [\aap] {10.1051/0004-6361/201322068}, \href {https://ui.adsabs.harvard.edu/#abs/2013A&A...558A..33A} {558, A33}

\bibitem[\protect\citeauthoryear{{Astropy Collaboration} et~al.,}{{Astropy Collaboration} et~al.}{2018}]{2018AstroPy}
{Astropy Collaboration} et~al., 2018, \mn@doi [\aj] {10.3847/1538-3881/aabc4f}, \href {http://adsabs.harvard.edu/abs/2018AJ....156..123A} {156, 123}

\bibitem[\protect\citeauthoryear{{Astropy Collaboration} et~al.,}{{Astropy Collaboration} et~al.}{2022}]{2022AstroPy}
{Astropy Collaboration} et~al., 2022, \mn@doi [\apj] {10.3847/1538-4357/ac7c74}, \href {https://ui.adsabs.harvard.edu/abs/2022ApJ...935..167A} {935, 167}

\bibitem[\protect\citeauthoryear{{Augereau} \& {Beust}}{{Augereau} \& {Beust}}{2006}]{2006AB}
{Augereau} J.~C.,  {Beust} H.,  2006, \mn@doi [\aap] {10.1051/0004-6361:20054250}, \href {https://ui.adsabs.harvard.edu/abs/2006A&A...455..987A} {455, 987}

\bibitem[\protect\citeauthoryear{{Babusiaux} et~al.,}{{Babusiaux} et~al.}{2023}]{2023bGaia}
{Babusiaux} C.,  et~al., 2023, \mn@doi [\aap] {10.1051/0004-6361/202243790}, \href {https://ui.adsabs.harvard.edu/abs/2023A&A...674A..32B} {674, A32}

\bibitem[\protect\citeauthoryear{{Beichman}, {Neugebauer}, {Habing}, {Clegg}  \& {Chester}}{{Beichman} et~al.}{1988}]{1988Beichman}
{Beichman} C.~A.,  {Neugebauer} G.,  {Habing} H.~J.,  {Clegg} P.~E.,   {Chester} T.~J.,  eds, 1988, {Infrared Astronomical Satellite (IRAS) Catalogs and Atlases.Volume 1: Explanatory Supplement.} ~1 Vol. 1

\bibitem[\protect\citeauthoryear{{Beuzit} et~al.,}{{Beuzit} et~al.}{2019}]{2019Beuzit}
{Beuzit} J.~L.,  et~al., 2019, \mn@doi [\aap] {10.1051/0004-6361/201935251}, \href {https://ui.adsabs.harvard.edu/abs/2019A&A...631A.155B} {631, A155}

\bibitem[\protect\citeauthoryear{{Bohren} \& {Huffman}}{{Bohren} \& {Huffman}}{1998}]{1998BohrenHuffman}
{Bohren} C.~F.,  {Huffman} D.~R.,  1998, {Absorption and Scattering of Light by Small Particles}.
{John Wiley \& Sons}

\bibitem[\protect\citeauthoryear{{Booth} et~al.,}{{Booth} et~al.}{2024}]{2024Booth}
{Booth} M.,  et~al., 2024, \mn@doi [arXiv e-prints] {10.48550/arXiv.2407.01413}, \href {https://ui.adsabs.harvard.edu/abs/2024arXiv240701413B} {p. arXiv:2407.01413}

\bibitem[\protect\citeauthoryear{{Bressan}, {Marigo}, {Girardi}, {Salasnich}, {Dal Cero}, {Rubele}  \& {Nanni}}{{Bressan} et~al.}{2012}]{bressan2012}
{Bressan} A.,  {Marigo} P.,  {Girardi} L.,  {Salasnich} B.,  {Dal Cero} C.,  {Rubele} S.,   {Nanni} A.,  2012, \mn@doi [\mnras] {10.1111/j.1365-2966.2012.21948.x}, \href {https://ui.adsabs.harvard.edu/abs/2012MNRAS.427..127B} {427, 127}

\bibitem[\protect\citeauthoryear{Burnham \& Anderson}{Burnham \& Anderson}{2004}]{2004Burnham}
Burnham K.~P.,  Anderson D.~R.,  2004, \mn@doi [Sociological Methods \& Research] {10.1177/0049124104268644}, 33, 261

\bibitem[\protect\citeauthoryear{{Burns}, {Lamy}  \& {Soter}}{{Burns} et~al.}{1979}]{1979BurnsLamySoter}
{Burns} J.~A.,  {Lamy} P.~L.,   {Soter} S.,  1979, \mn@doi [\icarus] {10.1016/0019-1035(79)90050-2}, \href {http://adsabs.harvard.edu/abs/1979Icar...40....1B} {40, 1}

\bibitem[\protect\citeauthoryear{{Chen}, {Mittal}, {Kuchner}, {Forrest}, {Lisse}, {Manoj}, {Sargent}  \& {Watson}}{{Chen} et~al.}{2014a}]{2014Chen}
{Chen} C.~H.,  {Mittal} T.,  {Kuchner} M.,  {Forrest} W.~J.,  {Lisse} C.~M.,  {Manoj} P.,  {Sargent} B.~A.,   {Watson} D.~M.,  2014a, \mn@doi [\apjs] {10.1088/0067-0049/211/2/25}, \href {https://ui.adsabs.harvard.edu/abs/2014ApJS..211...25C} {211, 25}

\bibitem[\protect\citeauthoryear{{Chen}, {Girardi}, {Bressan}, {Marigo}, {Barbieri}  \& {Kong}}{{Chen} et~al.}{2014b}]{chen2014}
{Chen} Y.,  {Girardi} L.,  {Bressan} A.,  {Marigo} P.,  {Barbieri} M.,   {Kong} X.,  2014b, \mn@doi [\mnras] {10.1093/mnras/stu1605}, \href {https://ui.adsabs.harvard.edu/abs/2014MNRAS.444.2525C} {444, 2525}

\bibitem[\protect\citeauthoryear{{Chen}, {Bressan}, {Girardi}, {Marigo}, {Kong}  \& {Lanza}}{{Chen} et~al.}{2015}]{chen2015}
{Chen} Y.,  {Bressan} A.,  {Girardi} L.,  {Marigo} P.,  {Kong} X.,   {Lanza} A.,  2015, \mn@doi [\mnras] {10.1093/mnras/stv1281}, \href {https://ui.adsabs.harvard.edu/abs/2015MNRAS.452.1068C} {452, 1068}

\bibitem[\protect\citeauthoryear{{Choquet} et~al.,}{{Choquet} et~al.}{2017}]{2017Choquet}
{Choquet} {\'E}.,  et~al., 2017, \mn@doi [\apjl] {10.3847/2041-8213/834/2/L12}, \href {http://adsabs.harvard.edu/abs/2017ApJ...834L..12C} {834, L12}

\bibitem[\protect\citeauthoryear{{Choquet} et~al.,}{{Choquet} et~al.}{2018}]{2018Choquet}
{Choquet} {\'E}.,  et~al., 2018, \mn@doi [\apj] {10.3847/1538-4357/aaa892}, \href {https://ui.adsabs.harvard.edu/abs/2018ApJ...854...53C} {854, 53}

\bibitem[\protect\citeauthoryear{{Cronin-Coltsmann} et~al.,}{{Cronin-Coltsmann} et~al.}{2021}]{2021Cronin-Coltsmann}
{Cronin-Coltsmann} P.~F.,  et~al., 2021, \mn@doi [\mnras] {10.1093/mnras/stab1237}, \href {https://ui.adsabs.harvard.edu/abs/2021MNRAS.504.4497C} {504, 4497}

\bibitem[\protect\citeauthoryear{{Cronin-Coltsmann}, {Kennedy}, {Kral}, {Lestrade}, {Marino}, {Matr{\`a}}  \& {Wyatt}}{{Cronin-Coltsmann} et~al.}{2023}]{2023Cronin}
{Cronin-Coltsmann} P.~F.,  {Kennedy} G.~M.,  {Kral} Q.,  {Lestrade} J.-F.,  {Marino} S.,  {Matr{\`a}} L.,   {Wyatt} M.~C.,  2023, \mn@doi [\mnras] {10.1093/mnras/stad3083}, \href {https://ui.adsabs.harvard.edu/abs/2023MNRAS.526.5401C} {526, 5401}

\bibitem[\protect\citeauthoryear{{Crotts} et~al.,}{{Crotts} et~al.}{2024}]{2024Crotts}
{Crotts} K.~A.,  et~al., 2024, \mn@doi [\apj] {10.3847/1538-4357/ad0e69}, \href {https://ui.adsabs.harvard.edu/abs/2024ApJ...961..245C} {961, 245}

\bibitem[\protect\citeauthoryear{{Dahlqvist} et~al.,}{{Dahlqvist} et~al.}{2022}]{2022Dahlqvist}
{Dahlqvist} C.~H.,  et~al., 2022, \mn@doi [\aap] {10.1051/0004-6361/202244145}, \href {https://ui.adsabs.harvard.edu/abs/2022A&A...666A..33D} {666, A33}

\bibitem[\protect\citeauthoryear{{Daley} et~al.,}{{Daley} et~al.}{2019}]{2019Daley}
{Daley} C.,  et~al., 2019, \mn@doi [\apj] {10.3847/1538-4357/ab1074}, \href {https://ui.adsabs.harvard.edu/abs/2019ApJ...875...87D} {875, 87}

\bibitem[\protect\citeauthoryear{{David} \& {Hillenbrand}}{{David} \& {Hillenbrand}}{2015}]{2015Trevor}
{David} T.~J.,  {Hillenbrand} L.~A.,  2015, \mn@doi [\apj] {10.1088/0004-637X/804/2/146}, \href {https://ui.adsabs.harvard.edu/abs/2015ApJ...804..146D} {804, 146}

\bibitem[\protect\citeauthoryear{{Dohnanyi}}{{Dohnanyi}}{1969}]{1969Dohnanyi}
{Dohnanyi} J.~S.,  1969, \mn@doi [\jgr] {10.1029/JB074i010p02531}, \href {http://adsabs.harvard.edu/abs/1969JGR....74.2531D} {74, 2531}

\bibitem[\protect\citeauthoryear{{Draine}}{{Draine}}{2003}]{2003Draine}
{Draine} B.~T.,  2003, \mn@doi [\apj] {10.1086/379118}, \href {http://adsabs.harvard.edu/abs/2003ApJ...598.1017D} {598, 1017}

\bibitem[\protect\citeauthoryear{{Dullemond}, {Juhasz}, {Pohl}, {Sereshti}, {Shetty}, {Peters}, {Commercon}  \& {Flock}}{{Dullemond} et~al.}{2012}]{2012Dullemond}
{Dullemond} C.~P.,  {Juhasz} A.,  {Pohl} A.,  {Sereshti} F.,  {Shetty} R.,  {Peters} T.,  {Commercon} B.,   {Flock} M.,  2012, {RADMC-3D: A multi-purpose radiative transfer tool}, Astrophysics Source Code Library, record ascl:1202.015

\bibitem[\protect\citeauthoryear{{Eiroa} et~al.,}{{Eiroa} et~al.}{2013}]{2013Eiroa}
{Eiroa} C.,  et~al., 2013, \mn@doi [\aap] {10.1051/0004-6361/201321050}, \href {https://ui.adsabs.harvard.edu/abs/2013A&A...555A..11E} {555, A11}

\bibitem[\protect\citeauthoryear{{Esposito} et~al.,}{{Esposito} et~al.}{2020}]{2020Esposito}
{Esposito} T.~M.,  et~al., 2020, \mn@doi [\aj] {10.3847/1538-3881/ab9199}, \href {https://ui.adsabs.harvard.edu/abs/2020AJ....160...24E} {160, 24}

\bibitem[\protect\citeauthoryear{{Faramaz} et~al.,}{{Faramaz} et~al.}{2021}]{2021Faramaz}
{Faramaz} V.,  et~al., 2021, \mn@doi [\aj] {10.3847/1538-3881/abf4e0}, \href {https://ui.adsabs.harvard.edu/abs/2021AJ....161..271F} {161, 271}

\bibitem[\protect\citeauthoryear{Foreman-Mackey}{Foreman-Mackey}{2016}]{2016ForemanMackey}
Foreman-Mackey D.,  2016, \mn@doi [The Journal of Open Source Software] {10.21105/joss.00024}, 24

\bibitem[\protect\citeauthoryear{{Foreman-Mackey}, {Hogg}, {Lang}  \& {Goodman}}{{Foreman-Mackey} et~al.}{2013}]{2013ForemanMackey}
{Foreman-Mackey} D.,  {Hogg} D.~W.,  {Lang} D.,   {Goodman} J.,  2013, \mn@doi [\pasp] {10.1086/670067}, \href {https://ui.adsabs.harvard.edu/abs/2013PASP..125..306F} {125, 306}

\bibitem[\protect\citeauthoryear{{Fouesneau}}{{Fouesneau}}{2022}]{PyPhot}
{Fouesneau} M.,  2022, "pyphot", \url {https://github.com/mfouesneau/pyphot}

\bibitem[\protect\citeauthoryear{{Gagn{\'e}} et~al.,}{{Gagn{\'e}} et~al.}{2018}]{2018Gagne}
{Gagn{\'e}} J.,  et~al., 2018, \mn@doi [\apj] {10.3847/1538-4357/aaae09}, \href {https://ui.adsabs.harvard.edu/abs/2018ApJ...856...23G} {856, 23}

\bibitem[\protect\citeauthoryear{{Gaia Collaboration} et~al.,}{{Gaia Collaboration} et~al.}{2016}]{2016Gaia}
{Gaia Collaboration} et~al., 2016, \mn@doi [\aap] {10.1051/0004-6361/201629272}, \href {https://ui.adsabs.harvard.edu/abs/2016A&A...595A...1G} {595, A1}

\bibitem[\protect\citeauthoryear{{Gaia Collaboration} et~al.,}{{Gaia Collaboration} et~al.}{2023}]{2023aGaia}
{Gaia Collaboration} et~al., 2023, \mn@doi [\aap] {10.1051/0004-6361/202243940}, \href {https://ui.adsabs.harvard.edu/abs/2023A&A...674A...1G} {674, A1}

\bibitem[\protect\citeauthoryear{Harris et~al.,}{Harris et~al.}{2020}]{2020Harris}
Harris C.~R.,  et~al., 2020, \mn@doi [Nature] {10.1038/s41586-020-2649-2}, 585, 357

\bibitem[\protect\citeauthoryear{{Heras}, {Eiroa}, {del Burgo}, {Marshall}  \& {Montesinos}}{{Heras} et~al.}{2025}]{2025Heras}
{Heras} A.~M.,  {Eiroa} C.,  {del Burgo} C.,  {Marshall} J.~P.,   {Montesinos} B.,  2025, \mn@doi [\aap] {10.1051/0004-6361/202449826}, \href {https://ui.adsabs.harvard.edu/abs/2025A&A...694A.325H} {694, A325}

\bibitem[\protect\citeauthoryear{{H{\o}g} et~al.,}{{H{\o}g} et~al.}{2000}]{2000Hog}
{H{\o}g} E.,  et~al., 2000, \aap, \href {https://ui.adsabs.harvard.edu/abs/2000A&A...355L..27H} {355, L27}

\bibitem[\protect\citeauthoryear{{Holland} et~al.,}{{Holland} et~al.}{2017}]{2017Holland}
{Holland} W.~S.,  et~al., 2017, \mn@doi [\mnras] {10.1093/mnras/stx1378}, \href {https://ui.adsabs.harvard.edu/abs/2017MNRAS.470.3606H} {470, 3606}

\bibitem[\protect\citeauthoryear{{Horner} et~al.,}{{Horner} et~al.}{2020}]{2020Horner}
{Horner} J.,  et~al., 2020, \mn@doi [\pasp] {10.1088/1538-3873/ab8eb9}, \href {https://ui.adsabs.harvard.edu/abs/2020PASP..132j2001H} {132, 102001}

\bibitem[\protect\citeauthoryear{{Hudgins}, {Sandford}, {Allamandola}  \& {Tielens}}{{Hudgins} et~al.}{1993}]{1993Hudgins}
{Hudgins} D.~M.,  {Sandford} S.~A.,  {Allamandola} L.~J.,   {Tielens} A.~G.~G.~M.,  1993, \mn@doi [\apjs] {10.1086/191796}, \href {https://ui.adsabs.harvard.edu/abs/1993ApJS...86..713H} {86, 713}

\bibitem[\protect\citeauthoryear{{Hughes}, {Duch{\^e}ne}  \& {Matthews}}{{Hughes} et~al.}{2018}]{2018Hughes}
{Hughes} A.~M.,  {Duch{\^e}ne} G.,   {Matthews} B.~C.,  2018, \mn@doi [\araa] {10.1146/annurev-astro-081817-052035}, \href {https://ui.adsabs.harvard.edu/abs/2018ARA&A..56..541H} {56, 541}

\bibitem[\protect\citeauthoryear{{Hunter}}{{Hunter}}{2007}]{2007Hunter}
{Hunter} J.~D.,  2007, \mn@doi [Computing in Science and Engineering] {10.1109/MCSE.2007.55}, \href {https://ui.adsabs.harvard.edu/#abs/2007CSE.....9...90H} {9, 90}

\bibitem[\protect\citeauthoryear{{Ishihara} et~al.,}{{Ishihara} et~al.}{2010}]{2010Ishihara}
{Ishihara} D.,  et~al., 2010, \mn@doi [\aap] {10.1051/0004-6361/200913811}, \href {https://ui.adsabs.harvard.edu/abs/2010A&A...514A...1I} {514, A1}

\bibitem[\protect\citeauthoryear{{Kains}, {Wyatt}  \& {Greaves}}{{Kains} et~al.}{2011}]{2011Kains}
{Kains} N.,  {Wyatt} M.~C.,   {Greaves} J.~S.,  2011, \mn@doi [\mnras] {10.1111/j.1365-2966.2011.18566.x}, \href {https://ui.adsabs.harvard.edu/abs/2011MNRAS.414.2486K} {414, 2486}

\bibitem[\protect\citeauthoryear{{Kavelaars} et~al.,}{{Kavelaars} et~al.}{2009}]{2009Kavelaars}
{Kavelaars} J.~J.,  et~al., 2009, \mn@doi [\aj] {10.1088/0004-6256/137/6/4917}, \href {https://ui.adsabs.harvard.edu/abs/2009AJ....137.4917K} {137, 4917}

\bibitem[\protect\citeauthoryear{{Kennedy} \& {Wyatt}}{{Kennedy} \& {Wyatt}}{2013}]{2013Kennedy}
{Kennedy} G.~M.,  {Wyatt} M.~C.,  2013, \mn@doi [\mnras] {10.1093/mnras/stt900}, \href {https://ui.adsabs.harvard.edu/abs/2013MNRAS.433.2334K} {433, 2334}

\bibitem[\protect\citeauthoryear{{Kennedy} et~al.,}{{Kennedy} et~al.}{2018}]{2018Kennedy}
{Kennedy} G.~M.,  et~al., 2018, \mn@doi [\mnras] {10.1093/mnras/sty492}, \href {https://ui.adsabs.harvard.edu/abs/2018MNRAS.476.4584K} {476, 4584}

\bibitem[\protect\citeauthoryear{{Klaassen} et~al.,}{{Klaassen} et~al.}{2024}]{2024Klaassen}
{Klaassen} P.,  et~al., 2024, \mn@doi [Open Research Europe] {10.12688/openreseurope.17450.1}, \href {https://ui.adsabs.harvard.edu/abs/2024ORE.....4..112K} {4, 112}

\bibitem[\protect\citeauthoryear{{Kla{\v{c}}ka}, {Kocifaj}, {P{\'a}stor}  \& {Petr{\v{z}}ala}}{{Kla{\v{c}}ka} et~al.}{2007}]{2007Klacka}
{Kla{\v{c}}ka} J.,  {Kocifaj} M.,  {P{\'a}stor} P.,   {Petr{\v{z}}ala} J.,  2007, \mn@doi [\aap] {10.1051/0004-6361:20066132}, \href {https://ui.adsabs.harvard.edu/abs/2007A&A...464..127K} {464, 127}

\bibitem[\protect\citeauthoryear{{Kossacki}}{{Kossacki}}{2021}]{2021Kossacki}
{Kossacki} K.~J.,  2021, \mn@doi [\icarus] {10.1016/j.icarus.2021.114613}, \href {https://ui.adsabs.harvard.edu/abs/2021Icar..36814613K} {368, 114613}

\bibitem[\protect\citeauthoryear{{Kral}, {Matr{\`a}}, {Wyatt}  \& {Kennedy}}{{Kral} et~al.}{2017}]{2017Kral}
{Kral} Q.,  {Matr{\`a}} L.,  {Wyatt} M.~C.,   {Kennedy} G.~M.,  2017, \mn@doi [\mnras] {10.1093/mnras/stx730}, \href {https://ui.adsabs.harvard.edu/abs/2017MNRAS.469..521K} {469, 521}

\bibitem[\protect\citeauthoryear{{Krist} et~al.,}{{Krist} et~al.}{2010}]{2010Krist}
{Krist} J.~E.,  et~al., 2010, \mn@doi [\aj] {10.1088/0004-6256/140/4/1051}, \href {https://ui.adsabs.harvard.edu/abs/2010AJ....140.1051K} {140, 1051}

\bibitem[\protect\citeauthoryear{{Krivov}}{{Krivov}}{2010}]{2010Krivov}
{Krivov} A.~V.,  2010, \mn@doi [Research in Astronomy and Astrophysics] {10.1088/1674-4527/10/5/001}, \href {http://adsabs.harvard.edu/abs/2010RAA....10..383K} {10, 383}

\bibitem[\protect\citeauthoryear{{Krivov} \& {Booth}}{{Krivov} \& {Booth}}{2018}]{2018KrivovBooth}
{Krivov} A.~V.,  {Booth} M.,  2018, \mn@doi [\mnras] {10.1093/mnras/sty1607}, \href {https://ui.adsabs.harvard.edu/abs/2018MNRAS.479.3300K} {479, 3300}

\bibitem[\protect\citeauthoryear{{Krivov} \& {Wyatt}}{{Krivov} \& {Wyatt}}{2021}]{2021KrivovWyatt}
{Krivov} A.~V.,  {Wyatt} M.~C.,  2021, \mn@doi [\mnras] {10.1093/mnras/staa2385}, \href {https://ui.adsabs.harvard.edu/abs/2021MNRAS.500..718K} {500, 718}

\bibitem[\protect\citeauthoryear{{Krivov}, {M{\"u}ller}, {L{\"o}hne}  \& {Mutschke}}{{Krivov} et~al.}{2008}]{2008Krivov}
{Krivov} A.~V.,  {M{\"u}ller} S.,  {L{\"o}hne} T.,   {Mutschke} H.,  2008, \mn@doi [\apj] {10.1086/591507}, \href {http://adsabs.harvard.edu/abs/2008ApJ...687..608K} {687, 608}

\bibitem[\protect\citeauthoryear{{Kuchner} \& {Stark}}{{Kuchner} \& {Stark}}{2010}]{2010KuchnerStark}
{Kuchner} M.~J.,  {Stark} C.~C.,  2010, \mn@doi [\aj] {10.1088/0004-6256/140/4/1007}, \href {https://ui.adsabs.harvard.edu/abs/2010AJ....140.1007K} {140, 1007}

\bibitem[\protect\citeauthoryear{{Lebouteiller}, {Barry}, {Spoon}, {Bernard-Salas}, {Sloan}, {Houck}  \& {Weedman}}{{Lebouteiller} et~al.}{2011}]{2011Lebouteiller}
{Lebouteiller} V.,  {Barry} D.~J.,  {Spoon} H.~W.~W.,  {Bernard-Salas} J.,  {Sloan} G.~C.,  {Houck} J.~R.,   {Weedman} D.~W.,  2011, \mn@doi [\apjs] {10.1088/0067-0049/196/1/8}, \href {https://ui.adsabs.harvard.edu/abs/2011ApJS..196....8L} {196, 8}

\bibitem[\protect\citeauthoryear{{Lestrade}, {Wyatt}, {Bertoldi}, {Menten}  \& {Labaigt}}{{Lestrade} et~al.}{2009}]{2009Lestrade}
{Lestrade} J.~F.,  {Wyatt} M.~C.,  {Bertoldi} F.,  {Menten} K.~M.,   {Labaigt} G.,  2009, \mn@doi [\aap] {10.1051/0004-6361/200912306}, \href {https://ui.adsabs.harvard.edu/abs/2009A&A...506.1455L} {506, 1455}

\bibitem[\protect\citeauthoryear{{Liu}, {Wang}  \& {Jiang}}{{Liu} et~al.}{2014}]{2014Liu}
{Liu} Q.,  {Wang} T.,   {Jiang} P.,  2014, \mn@doi [\aj] {10.1088/0004-6256/148/1/3}, \href {https://ui.adsabs.harvard.edu/abs/2014AJ....148....3L} {148, 3}

\bibitem[\protect\citeauthoryear{{Luppe}, {Krivov}, {Booth}  \& {Lestrade}}{{Luppe} et~al.}{2020}]{2020Luppe}
{Luppe} P.,  {Krivov} A.~V.,  {Booth} M.,   {Lestrade} J.-F.,  2020, \mn@doi [\mnras] {10.1093/mnras/staa2608}, \href {https://ui.adsabs.harvard.edu/abs/2020MNRAS.499.3932L} {499, 3932}

\bibitem[\protect\citeauthoryear{{MacGregor} et~al.,}{{MacGregor} et~al.}{2016}]{2016Macgregor}
{MacGregor} M.~A.,  et~al., 2016, \mn@doi [\apj] {10.3847/0004-637X/823/2/79}, \href {https://ui.adsabs.harvard.edu/abs/2016ApJ...823...79M} {823, 79}

\bibitem[\protect\citeauthoryear{{Marino}}{{Marino}}{2021}]{2021Marino}
{Marino} S.,  2021, \mn@doi [\mnras] {10.1093/mnras/stab771}, \href {https://ui.adsabs.harvard.edu/abs/2021MNRAS.503.5100M} {503, 5100}

\bibitem[\protect\citeauthoryear{{Marshall} et~al.,}{{Marshall} et~al.}{2014}]{2014aMarshall}
{Marshall} J.~P.,  et~al., 2014, \mn@doi [\aap] {10.1051/0004-6361/201424517}, \href {https://ui.adsabs.harvard.edu/abs/2014A&A...570A.114M} {570, A114}

\bibitem[\protect\citeauthoryear{{Marshall}, {Maddison}, {Thilliez}, {Matthews}, {Wilner}, {Greaves}  \& {Holland}}{{Marshall} et~al.}{2017}]{2017Marshall}
{Marshall} J.~P.,  {Maddison} S.~T.,  {Thilliez} E.,  {Matthews} B.~C.,  {Wilner} D.~J.,  {Greaves} J.~S.,   {Holland} W.~S.,  2017, \mn@doi [\mnras] {10.1093/mnras/stx645}, \href {https://ui.adsabs.harvard.edu/abs/2017MNRAS.468.2719M} {468, 2719}

\bibitem[\protect\citeauthoryear{{Marshall}, {Milli}, {Choquet}, {del Burgo}, {Kennedy}, {Matr{\`a}}, {Ertel}  \& {Boccaletti}}{{Marshall} et~al.}{2018}]{2018Marshall}
{Marshall} J.~P.,  {Milli} J.,  {Choquet} {\'E}.,  {del Burgo} C.,  {Kennedy} G.~M.,  {Matr{\`a}} L.,  {Ertel} S.,   {Boccaletti} A.,  2018, \mn@doi [\apj] {10.3847/1538-4357/aaec6a}, \href {https://ui.adsabs.harvard.edu/abs/2018ApJ...869...10M} {869, 10}

\bibitem[\protect\citeauthoryear{{Marshall}, {Wang}, {Kennedy}, {Zeegers}  \& {Scicluna}}{{Marshall} et~al.}{2021}]{2021Marshall}
{Marshall} J.~P.,  {Wang} L.,  {Kennedy} G.~M.,  {Zeegers} S.~T.,   {Scicluna} P.,  2021, \mn@doi [\mnras] {10.1093/mnras/staa3917}, \href {https://ui.adsabs.harvard.edu/abs/2021MNRAS.501.6168M} {501, 6168}

\bibitem[\protect\citeauthoryear{{Marshall} et~al.,}{{Marshall} et~al.}{2023a}]{2023aMarshall}
{Marshall} J.~P.,  et~al., 2023a, \mn@doi [\mnras] {10.1093/mnras/stad913}, \href {https://ui.adsabs.harvard.edu/abs/2023MNRAS.521.5940M} {521, 5940}

\bibitem[\protect\citeauthoryear{{Marshall}, {Cotton}, {Bott}, {Bailey}, {Kedziora-Chudczer}  \& {Brown}}{{Marshall} et~al.}{2023b}]{2023bMarshall}
{Marshall} J.~P.,  {Cotton} D.~V.,  {Bott} K.,  {Bailey} J.,  {Kedziora-Chudczer} L.,   {Brown} E.~L.,  2023b, \mn@doi [\mnras] {10.1093/mnras/stad979}, \href {https://ui.adsabs.harvard.edu/abs/2023MNRAS.522.2777M} {522, 2777}

\bibitem[\protect\citeauthoryear{{Matr{\`a}} et~al.,}{{Matr{\`a}} et~al.}{2025}]{2025Matra}
{Matr{\`a}} L.,  et~al., 2025, \mn@doi [\aap] {10.1051/0004-6361/202451397}, \href {https://ui.adsabs.harvard.edu/abs/2025A&A...693A.151M} {693, A151}

\bibitem[\protect\citeauthoryear{{Matthews} et~al.,}{{Matthews} et~al.}{2018}]{2018Matthews}
{Matthews} E.,  et~al., 2018, \mn@doi [\mnras] {10.1093/mnras/sty1778}, \href {https://ui.adsabs.harvard.edu/abs/2018MNRAS.480.2757M} {480, 2757}

\bibitem[\protect\citeauthoryear{{Montesinos} et~al.,}{{Montesinos} et~al.}{2016}]{2016Montesinos}
{Montesinos} B.,  et~al., 2016, \mn@doi [\aap] {10.1051/0004-6361/201628329}, \href {https://ui.adsabs.harvard.edu/abs/2016A&A...593A..51M} {593, A51}

\bibitem[\protect\citeauthoryear{{Morales} et~al.,}{{Morales} et~al.}{2009}]{2009Morales}
{Morales} F.~Y.,  et~al., 2009, \mn@doi [\apj] {10.1088/0004-637X/699/2/1067}, \href {https://ui.adsabs.harvard.edu/abs/2009ApJ...699.1067M} {699, 1067}

\bibitem[\protect\citeauthoryear{{Morales}, {Bryden}, {Werner}  \& {Stapelfeldt}}{{Morales} et~al.}{2016}]{2016Morales}
{Morales} F.~Y.,  {Bryden} G.,  {Werner} M.~W.,   {Stapelfeldt} K.~R.,  2016, \mn@doi [\apj] {10.3847/0004-637X/831/1/97}, \href {https://ui.adsabs.harvard.edu/abs/2016ApJ...831...97M} {831, 97}

\bibitem[\protect\citeauthoryear{{Mu{\~n}oz-Guti{\'e}rrez}, {Marshall}  \& {Peimbert}}{{Mu{\~n}oz-Guti{\'e}rrez} et~al.}{2023}]{2023Munoz}
{Mu{\~n}oz-Guti{\'e}rrez} M.~A.,  {Marshall} J.~P.,   {Peimbert} A.,  2023, \mn@doi [\mnras] {10.1093/mnras/stad218}, \href {https://ui.adsabs.harvard.edu/abs/2023MNRAS.520.3218M} {520, 3218}

\bibitem[\protect\citeauthoryear{{Mulders}, {Min}, {Dominik}, {Debes}  \& {Schneider}}{{Mulders} et~al.}{2013}]{2013Mulders}
{Mulders} G.~D.,  {Min} M.,  {Dominik} C.,  {Debes} J.~H.,   {Schneider} G.,  2013, \mn@doi [\aap] {10.1051/0004-6361/201219522}, \href {https://ui.adsabs.harvard.edu/abs/2013A&A...549A.112M} {549, A112}

\bibitem[\protect\citeauthoryear{{M{\"u}ller}, {L{\"o}hne}  \& {Krivov}}{{M{\"u}ller} et~al.}{2010}]{2010Muller}
{M{\"u}ller} S.,  {L{\"o}hne} T.,   {Krivov} A.~V.,  2010, \mn@doi [\apj] {10.1088/0004-637X/708/2/1728}, \href {https://ui.adsabs.harvard.edu/abs/2010ApJ...708.1728M} {708, 1728}

\bibitem[\protect\citeauthoryear{{Mustill} \& {Wyatt}}{{Mustill} \& {Wyatt}}{2009}]{2009Mustill}
{Mustill} A.~J.,  {Wyatt} M.~C.,  2009, \mn@doi [\mnras] {10.1111/j.1365-2966.2009.15360.x}, \href {https://ui.adsabs.harvard.edu/abs/2009MNRAS.399.1403M} {399, 1403}

\bibitem[\protect\citeauthoryear{{Najita}, {Kenyon}  \& {Bromley}}{{Najita} et~al.}{2022}]{2022Najita}
{Najita} J.~R.,  {Kenyon} S.~J.,   {Bromley} B.~C.,  2022, \mn@doi [\apj] {10.3847/1538-4357/ac37b6}, \href {https://ui.adsabs.harvard.edu/abs/2022ApJ...925...45N} {925, 45}

\bibitem[\protect\citeauthoryear{{Norfolk} et~al.,}{{Norfolk} et~al.}{2021}]{2021Norfolk}
{Norfolk} B.~J.,  et~al., 2021, \mn@doi [\mnras] {10.1093/mnras/stab1901}, \href {https://ui.adsabs.harvard.edu/abs/2021MNRAS.507.3139N} {507, 3139}

\bibitem[\protect\citeauthoryear{{Pan} \& {Schlichting}}{{Pan} \& {Schlichting}}{2012}]{2012PanSchlichting}
{Pan} M.,  {Schlichting} H.~E.,  2012, \mn@doi [\apj] {10.1088/0004-637X/747/2/113}, \href {https://ui.adsabs.harvard.edu/abs/2012ApJ...747..113P} {747, 113}

\bibitem[\protect\citeauthoryear{{Patten} \& {Willson}}{{Patten} \& {Willson}}{1991}]{1991PattenWillson}
{Patten} B.~M.,  {Willson} L.~A.,  1991, \mn@doi [\aj] {10.1086/115879}, \href {https://ui.adsabs.harvard.edu/abs/1991AJ....102..323P} {102, 323}

\bibitem[\protect\citeauthoryear{{Pawellek}, {Krivov}, {Marshall}, {Montesinos}, {{\'A}brah{\'a}m}, {Mo{\'o}r}, {Bryden}  \& {Eiroa}}{{Pawellek} et~al.}{2014}]{2014Pawellek}
{Pawellek} N.,  {Krivov} A.~V.,  {Marshall} J.~P.,  {Montesinos} B.,  {{\'A}brah{\'a}m} P.,  {Mo{\'o}r} A.,  {Bryden} G.,   {Eiroa} C.,  2014, \mn@doi [\apj] {10.1088/0004-637X/792/1/65}, \href {https://ui.adsabs.harvard.edu/abs/2014ApJ...792...65P} {792, 65}

\bibitem[\protect\citeauthoryear{{Pearce} et~al.,}{{Pearce} et~al.}{2022}]{2022Pearce}
{Pearce} T.~D.,  et~al., 2022, \mn@doi [\aap] {10.1051/0004-6361/202142720}, \href {https://ui.adsabs.harvard.edu/abs/2022A&A...659A.135P} {659, A135}

\bibitem[\protect\citeauthoryear{{Petit} et~al.,}{{Petit} et~al.}{2011}]{2011Petit}
{Petit} J.~M.,  et~al., 2011, \mn@doi [\aj] {10.1088/0004-6256/142/4/131}, \href {https://ui.adsabs.harvard.edu/abs/2011AJ....142..131P} {142, 131}

\bibitem[\protect\citeauthoryear{{Poynting}}{{Poynting}}{1903}]{1903Poynting}
{Poynting} J.~H.,  1903, \mnras, \href {https://ui.adsabs.harvard.edu/abs/1903MNRAS..64A...1P} {64, 1}

\bibitem[\protect\citeauthoryear{{Preibisch}, {Ossenkopf}, {Yorke}  \& {Henning}}{{Preibisch} et~al.}{1993}]{1993Preibisch}
{Preibisch} T.,  {Ossenkopf} V.,  {Yorke} H.~W.,   {Henning} T.,  1993, \aap, \href {https://ui.adsabs.harvard.edu/abs/1993A&A...279..577P} {279, 577}

\bibitem[\protect\citeauthoryear{{Ren} et~al.,}{{Ren} et~al.}{2023}]{2023Ren}
{Ren} B.~B.,  et~al., 2023, \mn@doi [\aap] {10.1051/0004-6361/202245458}, \href {https://ui.adsabs.harvard.edu/abs/2023A&A...672A.114R} {672, A114}

\bibitem[\protect\citeauthoryear{{Rhee}, {Song}, {Zuckerman}  \& {McElwain}}{{Rhee} et~al.}{2007}]{2007Rhee}
{Rhee} J.~H.,  {Song} I.,  {Zuckerman} B.,   {McElwain} M.,  2007, \mn@doi [\apj] {10.1086/509912}, \href {https://ui.adsabs.harvard.edu/abs/2007ApJ...660.1556R} {660, 1556}

\bibitem[\protect\citeauthoryear{{Riello} et~al.,}{{Riello} et~al.}{2021}]{riello2021}
{Riello} M.,  et~al., 2021, \mn@doi [\aap] {10.1051/0004-6361/202039587}, \href {https://ui.adsabs.harvard.edu/abs/2021A&A...649A...3R} {649, A3}

\bibitem[\protect\citeauthoryear{{Robertson}}{{Robertson}}{1937}]{1937Robertson}
{Robertson} H.~P.,  1937, \mn@doi [\mnras] {10.1093/mnras/97.6.423}, \href {https://ui.adsabs.harvard.edu/abs/1937MNRAS..97..423R} {97, 423}

\bibitem[\protect\citeauthoryear{Schwarz}{Schwarz}{1978}]{1978Schwarz}
Schwarz G.,  1978, \mn@doi [The Annals of Statistics] {10.1214/aos/1176344136}, 6, 461

\bibitem[\protect\citeauthoryear{{Sibthorpe}, {Kennedy}, {Wyatt}, {Lestrade}, {Greaves}, {Matthews}  \& {Duch{\^e}ne}}{{Sibthorpe} et~al.}{2018}]{2018Sibthorpe}
{Sibthorpe} B.,  {Kennedy} G.~M.,  {Wyatt} M.~C.,  {Lestrade} J.~F.,  {Greaves} J.~S.,  {Matthews} B.~C.,   {Duch{\^e}ne} G.,  2018, \mn@doi [\mnras] {10.1093/mnras/stx3188}, \href {https://ui.adsabs.harvard.edu/abs/2018MNRAS.475.3046S} {475, 3046}

\bibitem[\protect\citeauthoryear{{Skrutskie} et~al.,}{{Skrutskie} et~al.}{2006}]{2006Skrutskie}
{Skrutskie} M.~F.,  et~al., 2006, \mn@doi [\aj] {10.1086/498708}, \href {https://ui.adsabs.harvard.edu/abs/2006AJ....131.1163S} {131, 1163}

\bibitem[\protect\citeauthoryear{{Tang}, {Bressan}, {Rosenfield}, {Slemer}, {Marigo}, {Girardi}  \& {Bianchi}}{{Tang} et~al.}{2014}]{tang2014}
{Tang} J.,  {Bressan} A.,  {Rosenfield} P.,  {Slemer} A.,  {Marigo} P.,  {Girardi} L.,   {Bianchi} L.,  2014, \mn@doi [\mnras] {10.1093/mnras/stu2029}, \href {https://ui.adsabs.harvard.edu/abs/2014MNRAS.445.4287T} {445, 4287}

\bibitem[\protect\citeauthoryear{{Tazzari}, {Beaujean}  \& {Testi}}{{Tazzari} et~al.}{2018}]{2018Tazzari}
{Tazzari} M.,  {Beaujean} F.,   {Testi} L.,  2018, \mn@doi [\mnras] {10.1093/mnras/sty409}, \href {https://ui.adsabs.harvard.edu/abs/2018MNRAS.476.4527T} {476, 4527}

\bibitem[\protect\citeauthoryear{{Thureau} et~al.,}{{Thureau} et~al.}{2014}]{2014Thureau}
{Thureau} N.~D.,  et~al., 2014, \mn@doi [\mnras] {10.1093/mnras/stu1864}, \href {https://ui.adsabs.harvard.edu/abs/2014MNRAS.445.2558T} {445, 2558}

\bibitem[\protect\citeauthoryear{Virtanen et~al.,}{Virtanen et~al.}{2020}]{2020Virtanen}
Virtanen P.,  et~al., 2020, \mn@doi [Nature Methods] {10.1038/s41592-019-0686-2}, \href {https://rdcu.be/b08Wh} {17, 261}

\bibitem[\protect\citeauthoryear{{Vitense}, {Krivov}, {Kobayashi}  \& {L{\"o}hne}}{{Vitense} et~al.}{2012}]{2012Vitense}
{Vitense} C.,  {Krivov} A.~V.,  {Kobayashi} H.,   {L{\"o}hne} T.,  2012, \mn@doi [\aap] {10.1051/0004-6361/201118551}, \href {https://ui.adsabs.harvard.edu/abs/2012A&A...540A..30V} {540, A30}

\bibitem[\protect\citeauthoryear{{Wahhaj} et~al.,}{{Wahhaj} et~al.}{2016}]{2016Wahhaj}
{Wahhaj} Z.,  et~al., 2016, \mn@doi [\aap] {10.1051/0004-6361/201321887}, \href {http://cdsads.u-strasbg.fr/abs/2016arXiv161105866W} {596, L4}

\bibitem[\protect\citeauthoryear{{Warren} \& {Brandt}}{{Warren} \& {Brandt}}{2008}]{2008WarrenBrandt}
{Warren} S.~G.,  {Brandt} R.~E.,  2008, \mn@doi [Journal of Geophysical Research (Atmospheres)] {10.1029/2007JD009744}, \href {https://ui.adsabs.harvard.edu/abs/2008JGRD..11314220W} {113, D14220}

\bibitem[\protect\citeauthoryear{{Wenger} et~al.,}{{Wenger} et~al.}{2000}]{2000Wenger}
{Wenger} M.,  et~al., 2000, \mn@doi [\aaps] {10.1051/aas:2000332}, \href {https://ui.adsabs.harvard.edu/abs/2000A&AS..143....9W} {143, 9}

\bibitem[\protect\citeauthoryear{{Wolf} \& {Hillenbrand}}{{Wolf} \& {Hillenbrand}}{2003}]{2003Wolf}
{Wolf} S.,  {Hillenbrand} L.~A.,  2003, \mn@doi [\apj] {10.1086/377638}, \href {http://adsabs.harvard.edu/abs/2003ApJ...596..603W} {596, 603}

\bibitem[\protect\citeauthoryear{{Wright} et~al.,}{{Wright} et~al.}{2010}]{2010Wright}
{Wright} E.~L.,  et~al., 2010, \mn@doi [\aj] {10.1088/0004-6256/140/6/1868}, \href {https://ui.adsabs.harvard.edu/abs/2010AJ....140.1868W} {140, 1868}

\bibitem[\protect\citeauthoryear{{Wright} et~al.,}{{Wright} et~al.}{2023}]{2023Wright}
{Wright} G.~S.,  et~al., 2023, \mn@doi [\pasp] {10.1088/1538-3873/acbe66}, \href {https://ui.adsabs.harvard.edu/abs/2023PASP..135d8003W} {135, 048003}

\bibitem[\protect\citeauthoryear{{Wyatt}}{{Wyatt}}{2008a}]{2008WyattSmallBodies}
{Wyatt} M.~C.,  2008a, \mn@doi [arXiv e-prints] {10.48550/arXiv.0807.1272}, \href {https://ui.adsabs.harvard.edu/abs/2008arXiv0807.1272W} {p. arXiv:0807.1272}

\bibitem[\protect\citeauthoryear{{Wyatt}}{{Wyatt}}{2008b}]{2008Wyatt}
{Wyatt} M.~C.,  2008b, \mn@doi [\araa] {10.1146/annurev.astro.45.051806.110525}, \href {https://ui.adsabs.harvard.edu/abs/2008ARA&A..46..339W} {46, 339}

\bibitem[\protect\citeauthoryear{{Wyatt}, {Smith}, {Su}, {Rieke}, {Greaves}, {Beichman}  \& {Bryden}}{{Wyatt} et~al.}{2007}]{2007Wyatt}
{Wyatt} M.~C.,  {Smith} R.,  {Su} K.~Y.~L.,  {Rieke} G.~H.,  {Greaves} J.~S.,  {Beichman} C.~A.,   {Bryden} G.,  2007, \mn@doi [\apj] {10.1086/518404}, \href {https://ui.adsabs.harvard.edu/abs/2007ApJ...663..365W} {663, 365}

\bibitem[\protect\citeauthoryear{{Zhang}, {Green}  \& {Rix}}{{Zhang} et~al.}{2023}]{zhang2023}
{Zhang} X.,  {Green} G.~M.,   {Rix} H.-W.,  2023, \mn@doi [\mnras] {10.1093/mnras/stad1941}, \href {https://ui.adsabs.harvard.edu/abs/2023MNRAS.524.1855Z} {524, 1855}

\bibitem[\protect\citeauthoryear{{Zuckerman}}{{Zuckerman}}{2019}]{2019Zuckerman}
{Zuckerman} B.,  2019, \mn@doi [\apj] {10.3847/1538-4357/aaee66}, \href {https://ui.adsabs.harvard.edu/abs/2019ApJ...870...27Z} {870, 27}

\bibitem[\protect\citeauthoryear{{Zuckerman} \& {Song}}{{Zuckerman} \& {Song}}{2004}]{zuckerman2004}
{Zuckerman} B.,  {Song} I.,  2004, \mn@doi [\araa] {10.1146/annurev.astro.42.053102.134111}, \href {https://ui.adsabs.harvard.edu/abs/2004ARA&A..42..685Z} {42, 685}

\bibitem[\protect\citeauthoryear{{del Burgo} \& {Allende Prieto}}{{del Burgo} \& {Allende Prieto}}{2016}]{delburgo2016}
{del Burgo} C.,  {Allende Prieto} C.,  2016, \mn@doi [\mnras] {10.1093/mnras/stw2005}, \href {https://ui.adsabs.harvard.edu/abs/2016MNRAS.463.1400D} {463, 1400}

\bibitem[\protect\citeauthoryear{{del Burgo} \& {Allende Prieto}}{{del Burgo} \& {Allende Prieto}}{2018}]{delburgo2018}
{del Burgo} C.,  {Allende Prieto} C.,  2018, \mn@doi [\mnras] {10.1093/mnras/sty1371}, \href {https://ui.adsabs.harvard.edu/abs/2018MNRAS.479.1953D} {479, 1953}

\makeatother
\end{thebibliography}

\bsp	
\label{lastpage}







\appendix

\section{Dusty Grain Model}
\label{AppendixA}

A quasi-static dynamics and radiative transfer code for dust grains in debris discs (`Stardust') was developed to:

\begin{enumerate}
    \item predict the radial location of the dust grains depending on the forces acting on the grains due to the host star according to size (dynamics)
    \item predict the thermal emission of the dust grains from the grain’s location, size, and composition (radiative transfer)
\end{enumerate}

The grains in the Stardust model are assumed to be spherical objects that absorb and emit radiation according to Mie theory \citep{1998BohrenHuffman}.  In addition to gravity, these grains are subject to different forces that can move them far away from the point of production \citep[e.g.,][]{2008Krivov}.  The model can switch `on' and 'off' the force of stellar radiation pressure \citep[e.g.,][]{1979BurnsLamySoter} and the inward drag due to the Poynting-Robertson (P-R) effect \citep[e.g.,][]{1903Poynting,1937Robertson,2007Klacka} in comparison to the mean collisional time of the grain \citep[e.g.,][]{2010KuchnerStark}.  

The model assumes an optically thin disc with no gas present.  When the grains are traversing along elliptical orbits due to radiation pressure, it is assumed that no collisions occur. However, when the model activates the P-R effect, i.e. when the grain loses angular momentum and spirals inwards towards the star, a simple collision assumption is implemented based on the optical depth of the disc.  Additional force components, such as stellar wind, will be considered with future iterations of this model.

\begin{figure*}[t]
    \centering
    \includegraphics[width=\textwidth,trim = 3cm 1cm 3cm 0cm]{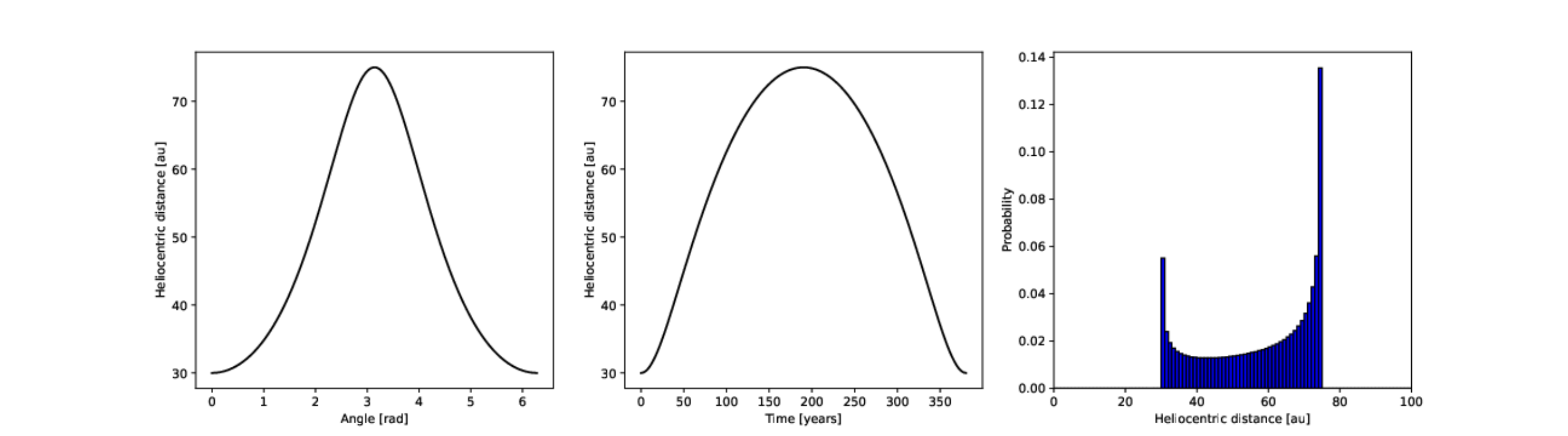}
    \caption{Grain corresponding to $\beta_{pr}$ = 0.3 being released at a stellar distance of 30 au. Left panel: stellar distance (au) of the grain as a function of True Anomaly (radians). Middle panel: stellar distance (au) of the grain as a function of time (years). Right panel: probability of finding the grain along its orbit as a function of stellar distance (au) binned in sizes of 1 au.}
    \label{fig:b0.3}
\end{figure*}

\subsection{Radiation pressure affected grains}

Grains in the model are initially on Keplerian orbits, where grains are under the influence of gravity. When the conservative radiation pressure force is switched on, this pushes grains out along elliptical orbits according to grain size. To determine the resultant orbit is by the ratio ($\beta$) of radiation pressure to gravitational forces \cite[e.g.,][]{2010Krivov}:

\begin{equation} \label{eq:beta_mod}
    \beta  = 0.574\frac{L_\star Q_{pr}(s)}{M_\star\rho s},
\end{equation}

where, $L_\star$ and $M_\star$ are the solar luminosity and mass of the host star, grain density ($\rho$) and size ($s$) are expressed in units of ‘g/cm$^3$’ and ‘$\micron$’ respectively, and the dimensionless quantity $Q_{pr}(s)$ is the radiation pressure efficiency dependent on grain size.  The $\beta$ values in the model are calculated for each corresponding grain size.

The eccentricity ($e$) of the orbit can be derived from the $\beta$ \citep[e.g.,][]{2006AB}:

\begin{equation}\label{eq:e-beta}
    e = \frac{\beta}{1-\beta}
\end{equation}

and the semi-major distance ($a$) of the ellipse, with pericentre ($r_p$), calculated using:

\begin{equation}\label{eq:a-e}
    a = \frac{r_p}{1-e}.
\end{equation}

Our aim is to determine the probability of a grain at a particular point in time along the orbit.  If we simply plot stellar distance as a function of True Anomaly ($\theta$), time is not considered (see left plot of Figure \ref{fig:b0.3} for a grain corresponding to {$\beta$~=~0.3}).  Therefore, to find a better representation we start with Kepler's Second Law in the following form:
\begin{equation} \label{eq:K2law}
    r^2d\theta = mabdt
\end{equation}

where $m = 2\pi/P$ is referred to as the mean motion, and $P$ is the orbital period in years for a given semi-major axis in au and host star mass in Solar masses ($P = a^{3/2}M_\star^{-1/2}$). The semi-major axis ($a$) and the semi-minor axis ($b$) are in units of au, and the period ($P$) is in years. Equation \ref{eq:K2law} cannot be solved analytically, thus, we have adopted the following iterative case:
\begin{equation}\label{eq:anglestep}
    d\theta = \Delta\theta = \theta_{n+1} - \theta_{n}
\end{equation} 

and the time-step defined by:
\begin{equation}\label{eq:timestep}
    dt = \Delta t~(= t_{n+1} - t_{n})
\end{equation}

So, substituting Equations \ref{eq:anglestep} and \ref{eq:timestep} into Equation \ref{eq:K2law}, yields the following relationship:

\begin{equation}\label{eq:angdisttime}
    \theta_{n+1} = \theta_{n} + \frac{abm}{r^{2}_{n}}\Delta t
\end{equation}

Equation \ref{eq:angdisttime} allows iterative solution to the stellar distance of the grain as a function of both True Anomaly and time. An appropriate timestep has been chosen to correspond to the $\beta$-value of the largest grain size. See middle plot of Figure \ref{fig:b0.3}, for the grain orbit corresponding to $\beta$ = 0.3. The next step is to group the radial data determined by solving Equation \ref{eq:angdisttime} with a bin size of 1~au (although in principle this can be any size desired), resulting in the relative probability distribution corresponding to $\beta$ in the right-hand plot of Figure \ref{fig:b0.3}. Inspecting this figure shows that the grain will spend most of its time on closest and farthest approaches of the star.  

This may be counter-intuitive for the nearest approach as the grain has the fastest speed. However, these extreme points (pericentre and apocentre) are turning points for the grain and the corresponding stellar distances ($r$) at these locales are more frequent, as opposed to locales where the grain is travelling towards and away from these extremities. 

\subsection{Grain size distribution}

The distribution of grain sizes typically favour smaller grains in debris discs.  \cite{1969Dohnanyi} presented a collisional model that showed the number density of particles per unit volume ($dn/dV$) in the mass range m to m + dm is:

\begin{equation}\label{eq:Dohnumden}
    \frac{dn}{dV} \propto m^{-\alpha}dm
\end{equation}

Where the exponent, $\alpha$, determines the rate of collisions.  For example, for $\alpha$ equal to 11/6 corresponds to a steady-state collisional cascade. However, in the literature \citep{1969Dohnanyi}, the steady-state collisional cascade model is often cited with the transform that shows the relative number of grains according to size:

\begin{equation} \label{eq:relnumsize}
    dN \propto s^{-q}ds
\end{equation}

Where $N$ is the number of grains corresponding to size $s$, and $q$ is the exponent that determines the type of collision. In this transform, a steady-state collisional cascade occurs when $q = 3.5$.  Generally, for debris discs, this exponent lies between 3.0 and 4.0 \citep{2018Hughes}.

\subsection{P-R Effect and collisions}

The model can include grains that undergo orbital decay due to loss of energy from the incident electromagnetic radiation on a moving non-rotating spherical grain, i.e. the P-R Effect. The time taken for grain to spiral in from outer orbit of semi-major axis ($a_i$) to an inner orbit of reduced semi-major axis ($a_f$) due to the P-R Effect \citep{2008WyattSmallBodies}:
\begin{equation}\label{eq:tpr}
    t_{\rm pr} = 400M^{-1}_{\star}(a_i^2-a_f^2)/{\beta},
\end{equation} 

For simplicity, the P-R time is compared to the (estimated) mean collisional time \citep[$t_{coll}$,][]{2010KuchnerStark}:

\begin{equation}\label{eq:tcoll}
    t_{coll} = \sqrt{(a_i^3/M)}/4\pi\tau,
\end{equation}

Where $\tau$ is the effective optical depth of planetesimal belt.

\subsection{Radiative transfer of grains}\label{s:radtrans}

We have described how the model defines grain location, size, and composition including belt structure.  We can now estimate the temperature and determine the direct thermal emission of the grains. 

If thermal equilibrium is assumed, i.e. the energy that the grains absorbs is equal to its emission, then the grain temperature ($T_g$) at a specific stellar distance ($r$) is derived from the following expression \citep{2003Wolf}:

\begin{equation} \label{eq:GraintTemp}
    r = \frac{R_\star}{2} \left[\frac{\int^{\infty}_0 Q_\lambda^{abs}(s)F_\lambda(T_\star) d\lambda}{\int^{\infty}_0 Q_\lambda^{abs}(s)B_\lambda(T_g) d\lambda}\right]^{1/2}
\end{equation}

where $R_\star$ and $T_\star$ are star's radius and effective temperature respectively. $F_\lambda(T)$ is the photosphere stellar model for star, $B_\lambda(T_g)$ is the blackbody function for a given grain temperature ($T_g$), and $Q_\lambda^{abs}(s)$ is the absorption efficiency calculated for a grain radius ($s$) as a function of wavelength. 

To determine the direct thermal emission, we will assume grains of a particular size ($s$) are within a rotationally symmetric disc heated by the host star observed from a distance $D$. These grains are within an annulus with mean stellar distance ($r$) with a thickness of one bin size. The direct thermal emission, measured in Janskys, is derived from a modified blackbody expression:

\begin{equation} \label{eq:fluxgrain}
    F^{th}_{\lambda}(r,s) = 4\pi^2N(r,s)\left(\frac{s}{D}\right)^2Q_\lambda^{abs}(s)B_{\lambda}(T_g),
\end{equation}

where the grain number density ($N(r,s)$) is defined as:

\begin{equation}
    N(r,s) =  2\pi n s r
\end{equation}

corresponding to the number of grains ($n$) of a specific grain size ($s$) that are present within an annulus with a mean radius defined by the stellar distance $r$.  It is noted that Equation \ref{eq:fluxgrain} is a modified form of the equation presented in both \cite{2008Krivov} and \cite{2010Muller} where they calculate the flux for all grains at all distances across the entire disc.

The optical properties of the Inclusion Matrix Particles (IMPs) were determined using Equation 8.50 of \cite{1998BohrenHuffman} where f = 0.5, 0.75, and 0.9 respectively; note this method was used in \cite{2016Morales}. These types of composition is commonly referred to as `dirty ice'.



\section{Posterior probability distributions of the modelling}
\label{AppendixB}
Here we present a corner plot showing the posterior probability distributions from the {\sc emcee} runs (10,0000 realisations of the model) and the values for the maximum likelihood model parameters presented in the text (Table \ref{tab:hd138965_alma_fit}, Section \ref{sec:mod}). We omit the posterior for the scale height parameter ($h$) from the corner plot as it is flat and uninformative.

\begin{figure*}
    \includegraphics[width=\textwidth]{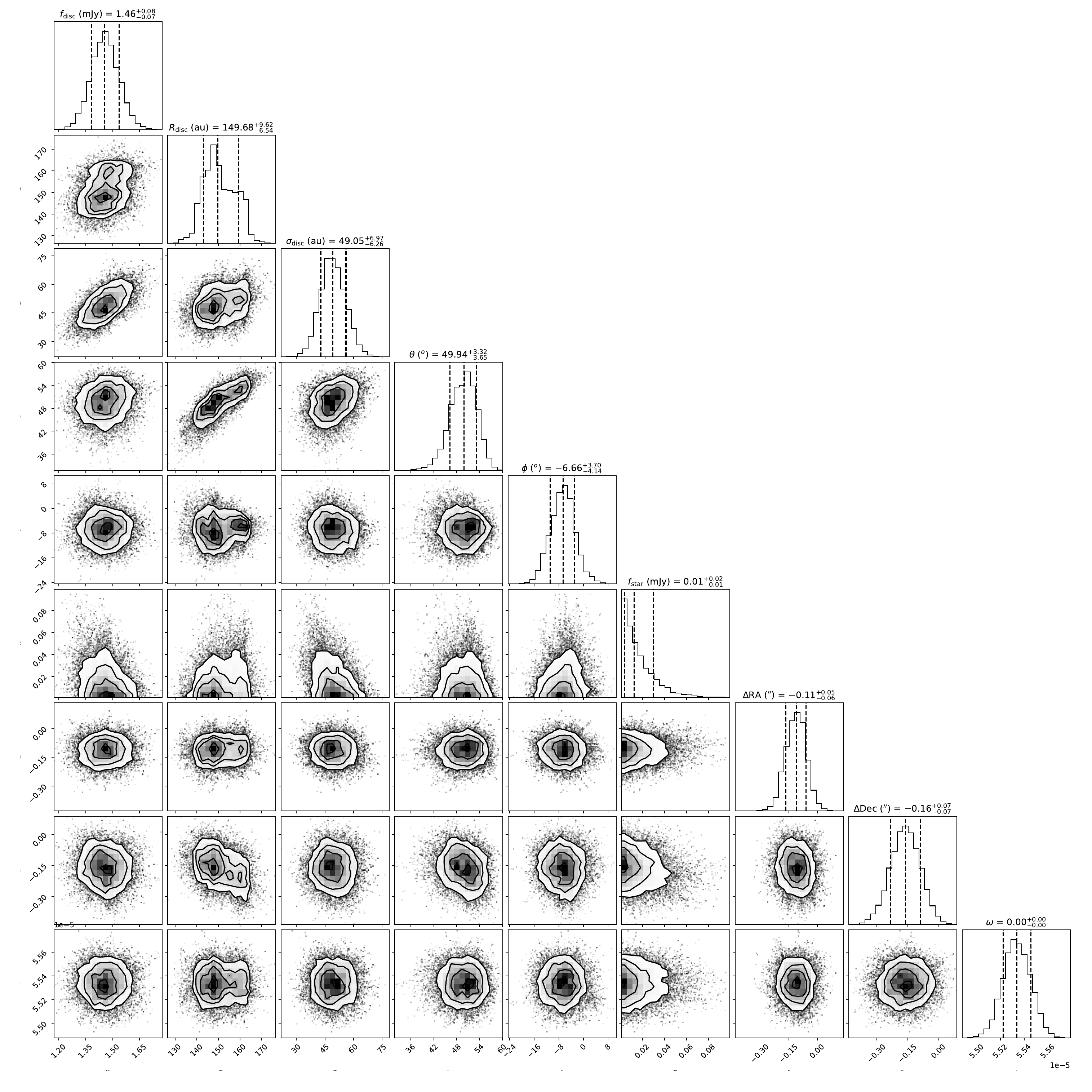}
    \caption{Corner plot showing the posterior probability distributions for the disc architecture model parameters. The posterior for the scale height, $h$, has been omitted as it is flat and uninformative. \label{fig:hd138965_corner_plot}}
\end{figure*}

\begin{figure*}\label{Fig:Corner_AS100}
    \includegraphics[width=\textwidth]{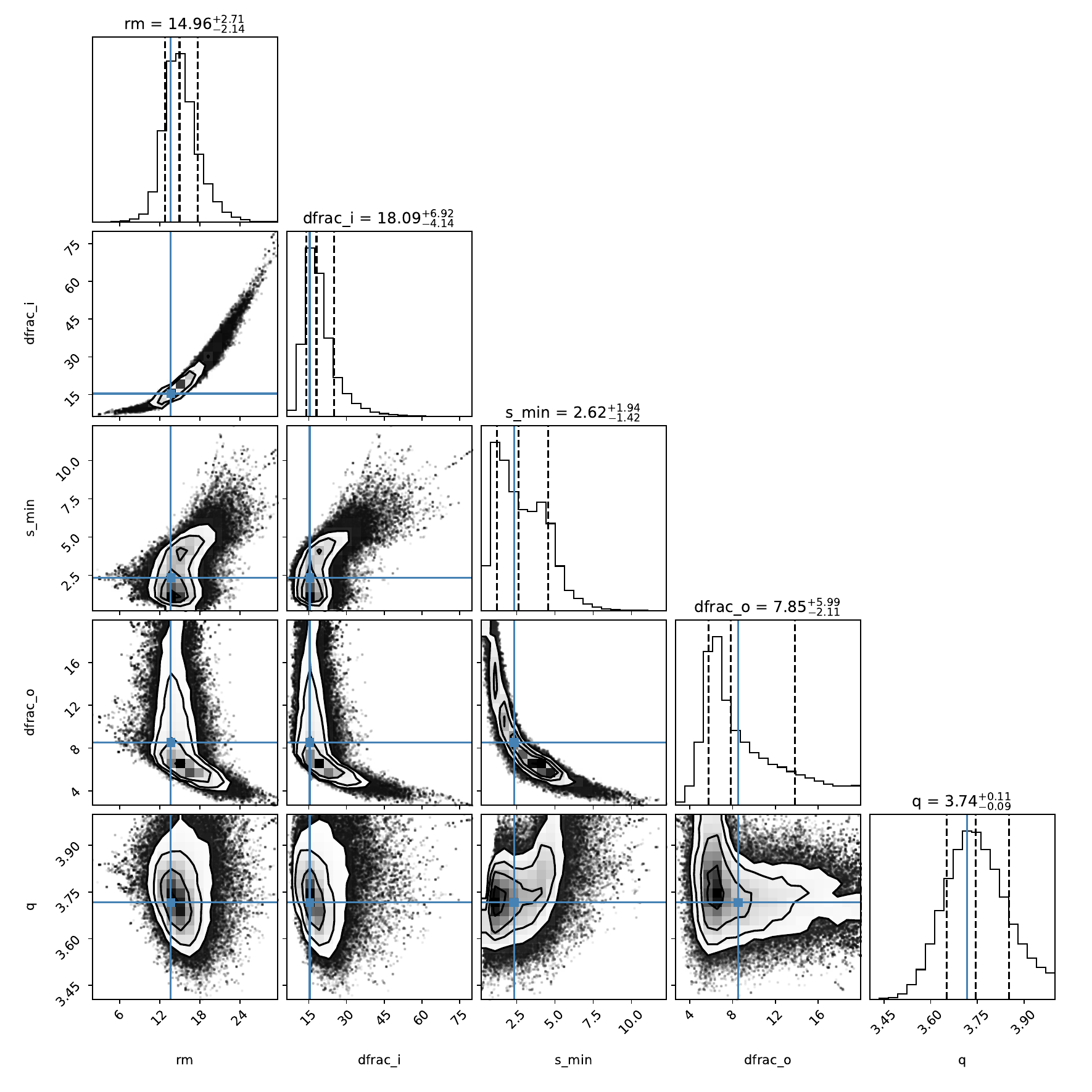}
    \caption{Corner plot showing the posterior probability distributions from the SED model analysis.  The blue `cross-hairs' are the Maximum Likelihood Values from the analysis. \label{fig:hd138965_SED_corner_plot}}
\end{figure*}

Here we also present the corner plot (Figure \ref{Fig:Corner_AS100})corresponding to the probability distribution to \textbf{to help pin down inner belt characteristics where the outer belt is composed of pure astronomical silicate.}

\end{document}